\documentstyle[12pt]{article}
\catcode`@=11
\def\@refe#1{#1}
\def\@biblabel#1{{\normalsize\bf{#1}}}
\def\refe{\@ifnextchar
[{\@tempswatrue\@citexr}{\@tempswafalse\@citexr[]}}
\def\@citexr[#1]#2{\if@filesw\immediate\write\@auxout{\string\citation
{#2}}\fi
  \def\@citea{}\@refe{\@for\@citeb:=#2\do
    {\@citea\def\@citea{,}\@ifundefined
       {b@\@citeb}{{\bf ?}\@warning
       {Citation `\@citeb' on page \thepage \space undefined}}%
\hbox{\csname b@\@citefb\endcsname}}}{#1}}
 \catcode`@=12
\title{ENTROPY OF RANDOM COVERINGS\\
AND\\
4-D QUANTUM GRAVITY}
\author{C.Bartocci \ddag, U.Bruzzo $\flat$, M.Carfora $\flat$,$\natural$,
 A.Marzuoli$\sharp$,$\natural$\\
\ddag\ Dipartimento di Matematica, Universit\'a di Genova\\
Via L.B. Alberti, 4 16132 Genova, Italy\\
$\flat$\ International School for Advanced Studies, SISSA-ISAS\\
Via Beirut 2-4, 34014 Trieste, Italy\\
$\natural$\ Istituto Nazionale di Fisica Nucleare, Sezione di
Pavia,Italy\\
$\sharp$\ Dipartimento di Fisica Nucleare e Teorica dell'Universit\`a
di
Pavia\\
Via Bassi 6, I-27100 Pavia, Italy}
\date{}
\def\Ricco{{\cal{R}}(n,r,D,V)}
\font\Bbbfont=msbm10
\def\Bbb#1{\hbox{\Bbbfont#1}}
\font\frakfont=eufm10
\def\frak#1{\hbox{\frakfont#1}}
\def\dim{\hbox{\rm dim}\,}
\def\Hom{\hbox{\rm Hom}\,}
\def\S{{\cal S}}\def\M{{\cal M}}
\def\unmezzo{\hbox{$1\over 2$}}
\def\op#1{\Omega^{#1}({\frak g}_{\theta})}

\begin{document}
\maketitle
\begin{abstract}
We discuss the counting of  minimal geodesic ball coverings of
$n$-dimensional  riemannian manifolds of bounded geometry,
fixed Euler characteristic and Reidemeister torsion in a
given representation of the fundamental group. This counting bears
relevance to the analysis of the
continuum limit of discrete models of quantum gravity.
We establish the conditions under which the number
of  coverings grows exponentially with the volume, thus
allowing for the
 search of a continuum limit of the corresponding discretized models.
The resulting entropy estimates depend on representations of the
fundamental group of the manifold through the corresponding
Reidemeister
torsion. We discuss the sum over inequivalent representations both
in the two-dimensional and in the four dimensional case.
Explicit entropy functions as well as significant bounds on the
associated critical
exponents are obtained in both cases.
\end{abstract}

\noindent {\bf SISSA Ref. 97/94/FM}\par \noindent
e-mail addresses:\par
\noindent
bartocci@matgen.dima.unige.it, bruzzo@sissa.it,\par
\noindent Carfora@pavia.infn.it,
Marzuoli@pavia.infn.it
\vfill\eject
\tableofcontents
\vfill\eject
\section{Introduction}

Dynamical triangulations, [ADF], [D2], [Ka], [We],
have recently  attracted much interest as
a computationally manageable method for the investigation of discrete
models of quantum gravity.
This approach  deals with a variant of Regge calculus [R], [Wi]
where, in alternative to the standard usage, the edge lengths  of
the triangulated manifolds
are kept fixed and set
equal to some minimal short-distance cut-off,
whereas
the underlying  combinatorial structure of the triangulations takes
the role of a statistical variable,
varying in some
ensemble of manifolds contributing to the model.
A dynamical content is thus given to the connectivity of
the triangulation in such a way that each choice of  a
triangulation corresponds to a choice of metric by Regge calculus.

 This particular
prominence given to the enumeration of triangulations  gives to
dynamically triangulated gravity the seemingly simple flavour of a
combinatorial theory. However,
it must be stressed that this simplicity is largely apparent rather
than actual, since
at a classical level and at a variance with standard Regge calculus,
dynamical triangulations do not afford a simple procedure for
recovering the Einstein-Hilbert action out of its
combinatorial counterpart,
diffeomorphism invariance being now completely lost. \par
\vskip 0.5 cm
The possible advantages in the use
of dynamical triangulations  are rather related to
the  different way in which one realizes, in this  approach, the
sampling of inequivalent Riemannian structures.
This is obtained by choosing a representative
metric, (by fixing the edge lengths), and by ergodically
varying the
combinatorial structure of the triangulation. We do not know of a
proof which explicitly shows a correspondence between this procedure
and a suitable continuous way of parametrizing the set of inequivalent
riemannian structures. Perhaps the Gromov-Hausdorff topology discussed
below provides such a correspondence. In any case, it is more or less
tacitly assumed that in this way one sweeps a much larger set of
riemannian structures as compared to the Regge case, where the
formalism, in this respect,
is less flexible owing to the constraints expressed  by the triangular
inequalities.
These constraints tend to localize the
edge-length varying triangulations used in Regge calculus  in a
neighborhood of the riemannian structure corresponding to the
triangulation originally given. Whereas,
one expects that the set of discretized manifolds considered in the
dynamically triangulated approach is uniformly distributed over the
space of all riemannian structures.\par
This is very
appealing for discussing the phase structure in the space of the
coupling constants of the theory: the cosmological constant, and the
gravitational coupling constant.
By defining the regularized
partition function as a  sum over topologically
equivalent triangulations, results  for continuum
quantum gravity can be extracted by looking for
critical points, in the space of  coupling constants, where the
observables of the model, such as the average number of simplexes,
diverge and obey scaling relations. This scaling behavior allows
for a renormalization of the couplings in terms of the given edge
length of the simplexes so as to obtain finite values for the volume
and other simple geometrical quantities characterizing the extended
configurations dominating the theory in the continuum limit. In other
words, one looks for the onset of a regime where the details of the
simplicial approximation become
irrelevant and a continuum theory can be constructed. \par
\vskip 0.5 cm
There is a  general comment that should be made at this
stage. In order to provide general entropy estimates
for discretized manifolds, we  find expedient to introduce another
kind of discretization, yet, besides dynamical triangulations and
Regge calculus. This discretization is associated with metric ball
coverings of given radius.  While not so useful from a numerical
point of view, it provides a good analytical edge on discrete quantum
gravity. It blends the simple combinatorial structure of dynamical
triangulations with the deep geometrical content of Regge calculus. We
feel that such variety of possible models should be considered with a
positive attitude, by taking advantage of the respective good
properties rather than emphasizing the drawbacks, as is often
done.   Thus, even if
in what follows we emphasize dynamical triangulations versus Regge
calculus, this does not mean that we wish to privilege that formalism
with respect to the other. The issue we address, the counting of the
number of topologically equivalent discretizations of an
$n$-manifold of given volume, ($n\geq 3$), is present in both cases (see
[Fro]), but it has been recently mostly emphasized for
dynamical triangulations.
\vskip 0.5 cm
As is well known, the main development of discrete models of quantum
gravity, and in particular of dynamically triangulated gravity, has
resulted from their role in providing a method for
regularizing non-critical bosonic string theory, (see {\it e.g.}
[FRS] for a review).
This latter can be seen as two-dimensional
quantum gravity interacting with $D$ scalar fields,
where $D$ is the dimension of the space where the string is embedded.
The  associated dynamically triangulated
models correctly reproduce, in the continuum limit, the results
obtained by conformal field theory. In particular, they are consistent
with the computation [KPZ], in the context of the Liouville model,
 of the entropy of closed surfaces
with Euler characteristic $\chi$, area $A$ and interacting with matter
fields with central charge $c\leq 1$, {\it viz.},
\begin{eqnarray}
S_{\chi}(A) \simeq
({\Lambda})^{A}
{A}^{\frac{{\chi}(\Sigma)}{2}({\gamma}_{str}-
2)-1}
\label{Superficie}
\end{eqnarray}
where  $\Lambda$ is a  suitable constant and ${\gamma}_{str}$,
the {\it string exponent}, is given
as a function of the central charge by
\begin{eqnarray}
{\gamma}_{str}=\frac{1}{12}(c-1-\sqrt{(25-c)(1-c)})
\end{eqnarray}
\vskip 0.5 cm
The above expression for ${\gamma}_{str}$ is valid as long as
$c\leq{1}$, and it appears to make sense only
in the weak coupling phase corresponding to
$c$, (or equivalently $D$), smaller than $1$.  For $c>1$,
conformal field theory becomes unstable,
and the above expression for the string exponent is no longer
reliable, (recently, an extension of the KZP scaling to the
$c>1$ case has been proposed by M.Martellini, M.Spreafico and K.
Yoshida,[MSY]).
Roughly speaking, it is believed that in this regime the
surfaces develop spikes and long tubes, and as seen from a
large distance the surface is no longer a two-dimensional
object. It collapses into a branched polymers configuration,
[DFJ].
It is important to stress that two-dimensional dynamically
triangulated models are well defined also in these cases,
where conformal field theory is no longer trustworthy, and they
provide a technique accessible to computer simulations.\par
\vskip 0.5 cm
A natural question concerns the possibility of extending the
techniques and some of the general results of the two-dimensional
case to the dimension three and four.
 This research program has been undertaken
by various groups by performing extensive computer simulations of
three- and four-dimensional triangulated manifolds.\par
Although
these simulated systems have a rather small size as compared to
the simulations used for 2D-gravity, (typically one puts together
$10^4$ four-simplexes, whereas in the two-dimensional case
triangulations with $10^7$ triangles are not unusual  [Ag]),
interesting results about critical phenomena
already emerge, (see [D1] for a an excellent review).
Such results are qualitatively similar in the $3D$ and $4D$ cases,
[Ag], [Aj], [Va],
in the sense that the phase diagram of the theory as a function of the
cosmological  constant and the gravitational coupling constant
shows the existence of a critical point. Here, the configurations
dominating the statistical sum change from being crumpled non-extended
objects to extended, finite Hausdorff-dimensional, objects. In three
dimensions there is a rather strong evidence that this change is
associated with a
first order transition indicating the absence of a continuum limit.
Whereas, in four dimensions  computer simulations indicate that the
transition between the crumpled and the extended phases may be of a
continuous nature.
\vskip 0.5 cm
There is increasing evidence to the soundness of this picture, and at
least from a general foundational point of view, dynamically
triangulated gravity seems to be now well established also in
dimension three and four. However, there still remain some outstanding
problems. The most obvious one is to obtain
explicit analytic
control on the theory,
(here we do not consider as dynamically triangulated models
the formulations of 3D-gravity \`a la Ponzano-Regge).
It is not yet not known if it is possible to obtain such a control,
and the best results at the moment come from
an interplay between computer simulations and the general analytic
properties of the various models considered, ({\it e.g.}, the choice
of the
most appropriate measure on the set of triangulated manifolds,
[BM]).\par
\vskip 0.5 cm
The experience with the two-dimensional case shows that
the delicate point here is to ascertain if the number of
dynamically triangulated $n$-manifolds, ($n>2$), of given
volume and fixed topology grows with the volume at most at an
exponential rate. This is a basic entropy bound necessary for having
the correct convergence properties of the partition function defining
the model.\par
\vskip 0.5 cm
In the case of surfaces, the required
entropy bounds, such as (\ref{Superficie}), are provided
either by direct counting
arguments, or by quantum field theory
techniques [BIZ], [FRS] as applied
to graph enumeration, a technique
that has found utility in a number of far reaching applications in
surface theory [Wt1], [Ko], [Pe].
In higher dimensions,  the natural generalizations of
such approaches are not viable even if
numerical as well as some  analytical
evidence [Am], [ADF], [Ag]
 shows that exponential bounds do hold in simple situations,
(typically for manifolds with $n$-sphere topology).
Recently, it has even been argued, on the basis of some numerical
evidence, that an exponential bound may fail to hold in dimension four
[CKR], but this analysis is quite controversial [AJ]. Conversely,
Boulatov has provided a nice argument for proving that for a
dynamically triangulated homotopy three-sphere there is an exponential
bound, [Bou], (the constants in the estimates are not characterized,
however).
Thus, a systematic
method for providing explicit entropic bounds relating
topology to the number of topologically equivalent triangulations
appears as a major open issue in higher
dimensional dynamically triangulated gravity [D1].
\vskip 0.5 cm
Without any control on the topology of manifolds,
there is no hope in
 the search for an exponentially bounded entropy function for the
number of  equivalent triangulations. For instance, it can be shown
[Am] that
the number of distinct triangulations on
(three)-manifolds,
with given
volume $V$ and arbitrary topological type, grows at least factorially
with $V$. Thus suitable constraints
on the class of riemannian
manifolds considered are necessary for having exponential growth
of the number of equivalent triangulations.\par
\noindent By analogy with the two-dimensional case, one may simply fix
the topology a priori, ({\it e.g.},
an $n$-sphere topology, $n=3$, $n=4$). This is a pragmatic point of
view. It has the
advantage of simplicity, but it has the serious drawback that it does
not allow to easily deal with fluctuating topologies,
 either because it is difficult to know
a priori what kind of topological invariants are going to enter the
entropy estimates in dimension $n\geq 3$, or because a topological
classification
of the relevant class of manifolds is often lacking, {\it e.g.},
in the case of three-manifolds.\par
\vskip 0.5 cm
The point of view implicit in the approach above is also motivated by
the assumption that
the topology of a manifold is not apparently under control in terms of
the geometrical invariants characterizing the size of a manifold, (and
hence its entropy), namely the volume or other simple geometrical
elements such as the diameter, and bounds on curvatures. \par
 However, the experience with recent developments in riemannian
geometry may suggest a change of this restrictive viewpoint.
Such an indication comes from a basic theorem due to Cheeger,
(see {\it e.g.}, [Ch] for a readable account of such finiteness
theorems),
according to which, for any given dimension, there are a {\it finite
number
of homeomorphism types} in the set of compact riemannian manifolds
with
volume bounded below, diameter bounded above and sectional curvature
bounded in absolute value. Further finiteness results of this type,
even under  weaker control on the size of the manifolds, have been
obtained [Pt], [GPW], recently.
A typical example in this sense is afforded by considering
for arbitrary $r\in {\bf R}$,
$v\in {\bf R}^+$, $D\in {\bf R}^+$ and integers $n\geq 3$,
the set  of closed connected Riemannian $n$-manifolds, $M$ whose
sectional curvatures satisfy $sec(M)\geq r$, whose volume satisfies
$Vol(M)\geq v$, and whose diameter is bounded above by
$D$, $diam(M)\leq D$. This is an
infinite-dimensional collection of riemannian structures, with
different underlying topologies. A huge space,
for which one can prove
finiteness of the homotopy types (in any dimension),
finiteness of the homeomorphism types (in dimension
$n=4$), and finiteness of diffeomorphism types (in any dimension
$n\geq 5$).\par
Even more generally, one may consider
the set of all metric spaces, (smooth manifolds,  and more general
spaces, {\it e.g.}, negatively curved polyhedra),
of Hausdorff dimension bounded above and for which a (Toponogov's)
comparison theorem for geodesic triangles locally holds, (Aleksandrov
spaces with curvature
bounded
below [BGP], [Per]). On a strict geometrical side, we wish to stress
that these are the spaces which  arise naturally if one wishes to
consider
simplicial approximations to riemannian manifolds.\par
\vskip 0.5 cm
It must be stressed that the imposition of (lower) bounds on sectional
curvatures does not seem to be fully consistent with the generic
triangulations considered in dynamically triangulated models of
quantum gravity. A simple two-dimensional example is afforded by
noticing that the local contribution to curvature corresponding to a
given vertex is ${\pi}/3(6-d)$, where $d$ is the order of the vertex,
({\it i.e.}, the number of edges meeting at it). A
priori, when considering dynamical triangulations, there is no natural
bound to the order $d$, and  the local curvature may grow arbitrarily
large. Thus spaces of bounded geometry may appear quite unsuitable as
an arena for discussing dynamically triangulated models.\par
The fact is that the use of spaces of bounded geometry should be
considered simply as a
technical step needed in order to get definite mathematical control on
problems raised when dealing with enumerative problems for
dynamical triangulations. In particular, once the entropy estimates
are obtained, we should  remove the dependence on the cutoffs
artificially introduced. A priori, this removal would call for a
rather delicate (inductive) limiting procedure,
{\it viz.}, considering the behaviour of the sequence of entropy
estimates on
the nested collection of spaces of bounded geometry obtained by
letting the lower bound to the curvature go to (minus) infinity.
Actually, the  entropy bounds obtained by us turn out to be
not sensible to the cutoffs, and the potential shortcomings of the use
of spaces of bounded geometry do not appear.
\vskip 0.5 cm
The possibility of getting some mathematical control on the entropy
problem by using spaces of bounded geometry is suggested by the
topological finiteness results
recalled above. To clarify somehow this
assertion, let us
recall that in any given dimension the set of manifolds which
satisfies the hypothesis of these finiteness theorems has
a compact closure in a Hausdorff-like topology [Gr1]. This
topology is naturally adapted to
the coarse grained point of view implicit in the discrete approaches
to quantum gravity, thus one may reasonably assume that  partition
functions associated with such discrete models are  continuous in such
topology.  Since the configuration
space is compact, and the partition functions are continuous, it follows
that out of the
sequence of bounded partition functions corresponding to finer and
finer
triangulations, we can extract a converging subsequence.
This implies the
corresponding existence of well-behaved entropy bounds.\par
\vskip 0.5 cm
Obviously, this is a heuristic argument, which however may serve as a
guiding principle. Indeed, following this viewpoint, we proved [CM4]
that, {\it up to a sum over inequivalent orthogonal representations of
the
fundamental group},
it is possible to explicitly provide the entropy function
counting the topologically equivalent ways of covering and
packing, with metric balls of given radius,
$n$-manifolds of bounded geometry, for any $n\geq 3$, (notice that
here topological equivalence stands for simple-homotopy equivalence).
Strictly
speaking this is not the entropy function for  dynamical
triangulations of the given manifold. However, it is easily seen, (see
the following paragraph),
that with a dynamically triangulated manifold  there is naturally
associated a metric ball covering, and that the number of
topologically equivalent metric ball coverings of given radius is
not-smaller  than the
number of corresponding dynamical triangulations. Thus, the entropy
function determined in [CM4] is an upper bound to the entropy function
for dynamical triangulations,
(for manifolds of bounded geometry). This argument is useful for
establishing that one has exponential bounds on the number of
equivalent triangulations. However, it is important to
stress that {\it it does not allow to determine the critical
exponents for dynamically triangulated models, (for $n\geq{3}$)}.
As a matter of fact, already for $n=2$, critical exponents for
geodesic ball coverings can be quite different from the
${\gamma}_{str}$ appearing in (\ref{Superficie}). This may be
seen as a rather obvious consequence of the intuitive fact
that there are many more states accessible to coverings
rather than to triangulations, since the latter are
combinatorially more rigid.

\vskip 0.5 cm

The analysis in [CM4] was rather incomplete, in particular
we did not attempt any
explicit determination of the critical exponents for geodesic
ball coverings, and the connection between this type of
discretization and the more familiar ones, like Regge calculus
and dynamical triangulations, was quite unclear.
Here we carry out an important step in this direction by
explicitly providing an
entropy estimate for geodesic ball coverings of
four-dimensional manifolds and by determining bounds to the
corresponding
critical exponent. On passing, we also discuss the two-dimensional
case, again by explicitly determining entropy estimates and bounds for
the
critical exponents.\par
\vskip 0.5 cm
\subsection{Summary of the results}

The results obtained can be summarized as follows.\par
\vskip 0.5 cm
\noindent {\it Entropy estimates in a given representation of the
fundamental group}\par
\vskip 0.5 cm
Let $M$ be a $n$-dimensional manifold, ($n\geq{2}$), of given
fundamental group ${\pi}_1(M)$, and let
$[\theta]\in Hom({\pi}_1(M),G)/G$ denote a conjugacy class of
representations of
${\pi}_1(M)$ into a Lie group endowed with an Ad-invariant, symmetric,
non-degenerate bilinear form, ({\it i.e.}, with
an Ad-invariant metric). \par
\vskip 0.5 cm
We think of $M$ as generated by a configuration
of $\lambda$ metric
balls, $\{B(i)\}$, of fixed radius $\epsilon$ in such a way that the
$\epsilon$-balls cover $M$ while the $\frac{\epsilon}{2}$-balls
are disjoint. Moreover, at most $d$ balls are allowed to mutually
overlap, (such $d$ depends on the geometry of the underlying manifold,
but it is otherwise independent from $\epsilon$).
We refer to the set of balls with radius $\epsilon/2$ as an
{\bf ${\epsilon}/2$-geodesic ball packing} of $M$, while the same
set of balls with radius $\epsilon$ defines the corresponding
{\bf $\epsilon$-geodesic ball covering} of $M$.\par
A priori, the balls are topologically non-trivial, namely both
the balls themselves and their mutual intersections are not
assumed to be contractible, (this allows for arbitrarily
large positive curvature in the underlying manifold). Explicitly,
the non-trivial topology of the balls is described by their
twisted cohomology groups $H^*_{\frak g}$ with coefficients
in a certain (adjoint) flat bundle associated with the representation
$\theta$. Roughly speaking, such groups provide {\it colours}
to the balls of the covering, and it is assumed that there are
$\lambda$ inequivalent colours to distribute
over the $\lambda$ balls. Any two such
colourings are considered
combinatorially inequivalent if the resulting patterns of the balls
belong to
distinct orbits of the action of the symmetric group acting
on the (centers of the) balls. We prove that such combinatorially
inequivalent
colourings can be used to construct, in the given representation
$\theta\colon{\pi}_1(M)\to{G}$, the {\it  distinct minimal geodesic
ball
coverings} of $M$, and thus, according to the previous remarks, they
can also be used to enumerate the number of topologically equivalent
triangulations, (definitions of what we mean for distinct coverings
and distinct dynamical triangulations of a given manifold $M$ are
given in section 2.1). \par
To be more precise on the meaning of topological equivalence adopted
here, it must be stressed that we are actually counting equivalent
triangulations having a common simple homotopy type.
This latter remark may need a few words of explanation.\par
\vskip 0.5 cm
A good counting function of utility for simplicial quantum gravity
should provide the number of geodesic ball coverings in manifolds
which are piecewise-linearly (PL) equivalent. But according to the
finiteness theorems recalled above, asking for such a counting
function is too much. In dimension three we have not yet control on
the enumeration of the homeomorphism types while in dimension four no
elementary enumeration is affordable for the PL types (by Cerf's
theorem we know that every PL 4-manifold carries a unique
differentiable structure; there can be only countably many
differentiable structures on a compact topological 4-manifolds, while
there are uncountably many diffeomorphism classes of 4-manifolds
homeomorphic to ${\Bbb R}^4$; in this sense  counting PL structures
is directly connected with the enumeration of differential structures).
Thus in the physically significant dimensions there is no obvious enumerative
criterion for PL structures.\par
The necessary compromise between what can be counted and what is of
utility for quantum gravity brings into evidence a particular
equivalence relation in homotopy known as {\it simple homotopy
equivalence}. Two polyhedra are simple-homotopy equivalent if they
have PL homeomorphic closed regular neighborhoods in some ${\Bbb
R}^n$. This notion of topological equivalence associated with simple
homotopy may seem too weak for our enumerative purposes, but as we
shall see it is sufficient for providing a detailed exponential bound
to the enumeration of dynamical triangulations.
\vskip 0.5 cm
It is also important to stress that even if the balls are
topologically
trivial, ({\it i.e.}, if they are contractible), the labelling
associated with the use of the twisted cohomology $H^*_{\frak g}$
is non-trivial. In such a case,  $H^*_{\frak g}$ reduces to
the assignment of the flat bundle, over the corresponding ball,
associated with the representation $\theta$. If all balls are
contractible, all such bundles are isomorphic, but, obviously,
not canonically. Thus,  $H^*_{\frak g}$ can be still
used as non trivial labels for counting purposes.\par
\vskip 0.5 cm

The explicit counting of the inequivalent orbits, under permutations
of the balls, associated with such colourings is obtained
by means of P\'olya's enumeration theorem, [Bo].
More precisely, P\'olya's theorem is used for counting geodesic
ball packings, so as to avoid the unwieldy complications arising
from the intersections of the balls when they cover the manifold.
The counting is then extended by a simple argument, (relying, however
on a deep compactness theorem by Gromov), to the geodesic ball
coverings associated with the packings.
\vskip 0.5 cm
{}From this enumeration
we get that, {\it in the given representation $\theta$}, the number,
$B_{Cov}({\Delta}^{\frak g},\lambda)$, of distinct geodesic ball
coverings with
$\lambda$ balls that can be
introduced in the manifold $M$
is bounded above, for large $\lambda$, by
\vskip 0.5 cm
\begin{eqnarray}
B_{Cov}({\Delta}^{\frak g},\lambda)\leq
\frac{1}{\sqrt{2\pi}{\Delta}^{\frak g}(M)}
\sqrt{\frac{n+2}{n+1}}
{\left [ \frac{(n+2)^{n+2}}{(n+1)^{n+1}}{\tilde w}
\right ] }^{\lambda}
{\lambda}^{-\frac{1}{2}}
\left( 1+O({\lambda}^{-\frac{3}{2}}) \right)
\label{summasi}
\end{eqnarray}
\vskip 0.5 cm
\noindent where $n$ denotes the dimension of $M$,
  ${\Delta}^{\frak g}(M)$ is the Reidemeister
torsion of $M$ in the {\it given representation}
$\theta\colon{\pi}_1(M)\to{G}$, and where ${\tilde w}$ is, roughly
speaking, the Reidemeister
torsion of the {\it dominant} twisted cohomology group of
the balls.  \par
\vskip 0.5 cm
Recall that, given a manifold $M$ and a representation of its
fundamental group ${\pi}_1(M)$ in a flat bundle
${\frak g}_{\theta}$, the Reidemeister torsion is a generalized
volume element constructed from the twisted cohomology groups
$H^i(M,{\frak g}_{\theta})$. In even dimension, if $M$ is compact,
orientable,
and without boundary, it can be shown by Poincar\'e duality that
${\Delta}^{\frak g}(M)=1$. However, this latter result does not
hold for the balls of the covering since they have a boundary.
In such a case, the corresponding torsion ${\tilde w}$ depends
non-trivially on the metric of the ball, too.
\vskip 0.5 cm
Topologically speaking, (\ref{summasi}) is estimating the number of
geodesic ball coverings on a manifold of given {\it
simple homotopy} type, (for a given ${\pi}_1(M)$ and a given
representation $\theta$, this simple homotopy type is characterized by
the torsion). If one
is interested in counting coverings, (and triangulations),
{\it just} on a manifold of given fundamental group, then
(\ref{summasi}) reduces to
\vskip 0.5 cm
\begin{eqnarray}
\frac{1}{\sqrt{2\pi}}
\sqrt{\frac{n+2}{n+1}}
{\left[ \frac{(n+2)^{n+2}}{(n+1)^{n+1}}\right] }^{\lambda}
{\lambda}^{-\frac{1}{2}}
\left( 1+O({\lambda}^{-\frac{3}{2}}) \right)
\end{eqnarray}
\vskip 0.5 cm
\noindent which does not depend any longer on the representation
$\theta\colon{\pi}_1(M)\to{G}$, and provides a significant exponential bound
to the number of distinct coverings that one can
introduce on $M$.
\vskip 0.5 cm
In particular, the number of distinct geodesic ball coverings, with
$\lambda$ balls, that can be introduced on a surface $\Sigma$ of given
topology turns out to be asymptotically bounded by
\vskip 0.5 cm
\begin{eqnarray}
\frac{2}{\sqrt{6\pi}}{\left [ \frac{4^4}{3^3} \right ]
}^{\lambda}{\lambda}^{-1/2}
\end{eqnarray}
\vskip 0.5 cm
This bound is perfectly consistent with the classical result provided
by W.Tutte [BIZ] according to which the number of distinct
triangulations, with $\lambda$ vertices, of a surface (with the
topology of the sphere) is asymptotically
\begin{eqnarray}
\frac{1}{64\sqrt{6\pi}}{\left [ \frac{4^4}{3^3} \right ]
}^{\lambda}{\lambda}^{-7/2}
\end{eqnarray}
\vskip 0.5 cm

The finer  entropy estimates (\ref{summasi}) do depend on the
particular
representation $\theta$, thus a more interesting object to discuss
is their average over all possible inequivalent representations
in the given group $G$ obtained by integrating (\ref{summasi}) over
the representation variety $Hom({\pi}_1(M),G)/G$.\par
\vskip 1 cm
\noindent {\it Entropy estimates at fixed $\lambda$, and $n=2$}\par
\vskip 0.5 cm
Denoting by $\theta$ the dominant representations, (in a formal saddle
point evaluation of the integral over inequivalent
representations), we get for the entropy estimate, up
to some inessential constants
\vskip 0.5 cm
\begin{eqnarray}
\int_{Hom({\pi}_1(M),G)/G}B_{Cov}({\Delta}^{\frak g},\lambda)
\leq\nonumber
\end{eqnarray}

\begin{eqnarray}
\sum_{\theta\in Hom_0}
\frac{2}{\sqrt{6\pi}{\Delta}_{\theta}^{\frak g}(M)}
{\left [ \frac{4^4}{3^3}{\tilde w}_{\theta}
\right ] }^{\lambda}
{\lambda}^{[-\frac{(h-1)}{2}dim(G)-
\frac{dim(z(\theta))}{2}-\frac{1}{2}]}
\left( 1+\ldots \right)
\label{summary1}
\end{eqnarray}
\vskip 0.5 cm
\noindent where $Hom_0$ denotes the (finite) set of representations
contributing to the saddle point evaluation,  $h$ denotes the
genus of the surface $M$, and  $z(\theta)$ denotes the
centralizer of ${\theta}({\pi}_1(M))$ in
the Lie group $G$.\par
\vskip 0.5 cm
We define the critical exponent ${\eta}(G)$ associated with the
entropy function  $B_{Cov}({\Delta}^{\frak g},\lambda)$
by means of the relation

\begin{eqnarray}
\int_{Hom({\pi}_1(M),G)/G}
B_{Cov}({\Delta}^{\frak g},\lambda)\equiv
Meas{\left(\frac{Hom({\pi}_1(M),G)}{G}\right) }
\exp[c\lambda]
{\lambda}^{{\eta}_{sup}-3}
\end{eqnarray}
\vskip 0.5 cm
\noindent where $c$ is a suitable constant, (depending on $G$).
Then (\ref{summary1}) provides also a bound for ${\eta}(G)$
given by (for a given $\theta\in Hom_0$),

\begin{eqnarray}
{\eta}(G)\leq 2+(1-h)\frac{dim(G)}{2}+\frac{1}{2}(1-dim(z(\theta)))
\label{duecritico}
\end{eqnarray}

For instance, for $G=U(1)$, we get

\begin{eqnarray}
{\eta}(G)\leq 2+\frac{1}{2}(1-h)
\end{eqnarray}

\noindent which is consistent with
KPZ scaling. This bound is an equality in the obvious case $h=1$,
while it is sharp in the remaining cases.
It is likely that  (\ref{duecritico}) holds
also in the case where there is a strong coupling of 2D-gravity with
matter, namely in the regime where KPZ scaling breaks down.
\vskip 1 cm
\noindent {\it Entropy estimates at fixed $\lambda$, and $n=4$}\par
\vskip 0.5 cm
In the four-dimensional case we obtain, again through a  formal saddle
point
evaluation, and up to some inessential factors

\begin{eqnarray}
\sum_{\theta\in Hom_0}
\frac{\sqrt{6}}{\sqrt{10\pi}{\Delta}_{\theta}^{\frak g}(M)}
{\left [ \frac{6^6}{5^5}{\tilde w}_{\theta}
\right ] }^{\lambda}
{\lambda}^{[dim(G){\chi}(M)/8-b(2)/8
-1/2]}
\left( 1+\ldots  \right)
\end{eqnarray}

 \noindent where ${\chi}(M)$ is the Euler-Poincar\'e characteristic of
$M$ and $b(2)$ is the second Betti number associated with $H^*_{\frak
g}(M)$. \par

\noindent  Notice  that in the above expressions
we can set ${\Delta}_{\theta}^{\frak g}({\cal O}_h)=1$, (the torsion
being trivial in  even dimensions for a closed, orientable manifold).
\vskip 0.5 cm
The bound on the critical
exponent corresponding to this entropy estimate
is (for a given $\theta\in Hom_0$),

\begin{eqnarray}
{\eta}(G)\leq\frac{5}{2}+
\frac{dim(G){\chi}(M)}{8}-\frac{b(2)}{8}
\end{eqnarray}
\vskip 0.5 cm

 This exponent, evaluated for the four-sphere, takes on the
value $\frac{11}{4}$ which is  larger than
the corresponding exponent obtained through numerical simulations,
(see {\it e.g.}, [Va]). In this latter case,
the available values of this exponent are typically affected by
a large uncertainty.
Nonetheless, numerical evidence seems
to indicate a rough value around the figures $0.40$, $0.57$,
thus  our bound is strict and likely not optimal. \vskip 0.5 cm

 We are perfectly aware that
this work is incomplete in many respects. In particular, it is
annoying that one does not get an entropy estimate directly for
triangulated four-manifolds but rather for
geodesic ball covered manifolds.  However, this estimate is sufficient
for controlling the number of topologically (in the simple-homotopical
sense) equivalent dynamical
triangulations on four-manifolds of bounded geometry, and it is, we
believe, a good starting point for a further understanding of discrete
models of four-dimensional quantum gravity.\par
\vskip 0.5 cm
We now turn to a more extensive discussion of our subject.\par
\vskip 0.5 cm
\section{Metric ball coverings and
triangulated manifolds }

As recalled in the introductory remarks, in order to regain a smooth
geometric perspective when dealing with a dynamically triangulated
manifold ${\cal T}$, we have to move our observation point far away
from
${\cal T}$, (for rather different reasons this same point of view,
which is the essence of a scaling limit, is advocated in geometric
group theory [Gr2] . In this
way, and under suitable re-scaling for the coupling constants of the
theory, the details of the triangulation ${\cal T}$ may fade away at
criticality, and the simplexes of
${\cal T}$ coalesce into  extended objects,  generalized metric
manifolds representing the {\it spacetime} manifolds (or more
correctly, an Euclidean version of them) dominating the statistical
sum of the model considered.\par
Technically speaking, this limiting procedure appeals here to a
topology in the set of metric spaces coming along with a
Hausdorff-type metric. This was rather explicitly suggested in
1981 by J.Fr\"{o}hlich [Fro] in his unpublished notes on Regge's
model. For completely different reasons, and more or less in the same
period, this notion of topology was made precise by M.Gromov [Gr1], and
used by him very effectively to discuss the compactness properties of
the space of riemannian structures. A detailed analysis is presented
in [Gr1],[CM2], and instead of repeating it here we give the intuition
and a few basic definitions. The rough idea is that
given a length cut-off $\epsilon$, two riemannian manifolds are to be
considered near in this topology, (one is the
$\epsilon$-Gromov-Haus\-dorff
approximation of the other), if their metric properties are
similar at length scales $L\geq{\epsilon}$. This intuition can be made
more precise as follows.\par
\vskip 0.5 cm
Consider two riemannian
   manifolds $M_1$   and $M_2$, (or more in general any two compact
metric spaces), let $d_{M_1}(\cdot,\cdot)$ and $d_{M_2}(\cdot,\cdot)$
respectively denote the corresponding distance functions, and let
$\phi\colon M_1\to M_2$ be a map between $M_1$ and $M_2$,
(this map is not required to be continuous ). If
$\phi$ is such that: {\it (i)}, the $\epsilon$-neighborhood of
${\phi}(M_1)$ in $M_2$ is equal to $M_2$, and {\it (ii)}, for
each $x$, $y$ in $M_1$ we have
\begin{eqnarray}
|d_{M_1}(x,y)-d_{M_2}({\phi}(x),{\phi}(y))|<\epsilon
\end{eqnarray}
then $\phi$ is said to be an $\epsilon$-{\it Hausdorff approximation}.
The  {\it Gromov-Hausdorff distance}
between the two
riemannian manifolds $M_1$ and  $M_2$,  $d_G(M_1,M_2)$, is then
defined according to [Gr1]
\newtheorem{Gaiad}{Definition}
\begin{Gaiad}
  $d_G(M_1,M_2)$ is the lower bound of the positive numbers $\epsilon$
such that there exist $\epsilon$-Hausdorff approximations from $M_1$
to $M_2$ and from $M_2$ to $M_1$.
\label{miauno}
\end{Gaiad}
\vskip 0.5 cm
The notion of $\epsilon$-Gromov-Haus\-dorff approximation is the weakest
large-scale equivalence relation between metric spaces of use in
geometry, and
is manifestly adapted to the needs of
simplicial quantum gravity, (think of a manifold and of a simplicial
approximation to it).\par
Notice that $d_G$ is not, properly
speaking,  a distance since it
does not satisfy the triangle inequality, but it rather gives rise to
a metrizable uniform structure in which the set of isometry classes
of all compact metric spaces, (not just riemannian structures),
is Hausdorff and complete. This enlarged space does
naturally contain topological (metric) manifolds and curved
polyhedra. As stressed in [Pt], the importance of this notion of
distance lies not so much in the fact that we have a distance
function, but in that we have a way of measuring when riemannian
manifolds, (or more general metric spaces), look alike.\par
\vskip 0.5 cm
In order to provide the entropy of four-dimensional triangulated
manifolds, we need to use Gromov-Hausdorff topology
quite superficially. Explicitly, it only appears in the ensemble of
manifolds for which we characterize the entropy function:
\begin{Gaiad}
For $r$ a real number, $D$ and $V$  positive real numbers, and
$n$ a natural number, let
$\Ricco$  denote the Gromov-Hausdorff closure of the space of isometry
classes of closed connected
 n-dimensional riemannian manifolds $(M,g)$ with
sectional curvature bounded below by $r$, {\it viz.},
\begin{eqnarray}
\inf_{x\in M}\{\inf \{ g_x(Riem_x(u,v)u;v) \colon u,v \in
T_xM,orthonormal\}\}\geq r  \nonumber
\end{eqnarray}
\noindent and diameter bounded above by $D$,
\begin{eqnarray}
diam(M)\equiv\sup_{(p,q)\in M\times M}d_M(p,q)\leq D\nonumber
\end{eqnarray}
\noindent and volume bounded below by $V$.
\label{miadue}
\end{Gaiad}
\vskip 0.5 cm

The point in the introduction of $\Ricco$ or of more general classes
of metric spaces with a lower bound on a suitably defined notion of
curvature, is that for any
manifold (or metric space) $M$ in such a class one gets a packing
information which is most helpful in controlling the topology in terms
of the metric geometry. In the case of $\Ricco$ this packing
information is provided by suitable coverings with geodesic (metric)
balls yielding
a coarse classification of the riemannian structures occurring in
$\Ricco$, (notice that these coverings can be introduced under
considerably less restrictive conditions, in particular it is
sufficient
to have a lower bound on the Ricci tensor, and an upper bound
on the diameter, [GP]).\par
\vskip 0.5 cm
In order to define such coverings [GP],
let us parametrize geodesics on $M\in \Ricco$ by arc length, and
for any $p\in M$ let us denote by ${\sigma}_p(x)\equiv d_M(x,p)$ the
distance function of the generic point $x$ from the chosen point $p$.
Recall that ${\sigma}_p(x)$ is a smooth function away from $\{p\cup
C_p\}$, where $C_p$, a closed nowhere dense set of measure zero, is
the cut locus of $p$. Recall also that a point $y\not= p$ is a
critical point of ${\sigma}_p(x)$ if for all vectors ${\bf v}\in
TM_y$, there is a minimal geodesic, $\gamma$,
from $y$ to $p$ such that the angle between ${\bf v}$ and
$\dot{\gamma}(0)$ is not greater than $\frac{\pi}{2}$.\par
\begin{Gaiad}
For any manifold $M\in \Ricco$  and for any
given  $\epsilon >0$, it is always
possible to find an ordered set of points $\{p_1,\ldots,p_N\}$ in $M$,
 so that, [GP]\par
\noindent  {\it (i)} the open metric balls, (the {\it geodesic
balls}),
$B_{M}(p_{i},\epsilon) = \{x \in M \vert d(x, p_{i})<
\epsilon\}$, $i=1,\ldots,N$, cover $M$; in
other words the collection
\begin{eqnarray}
{\{p_1,\ldots,p_N\}}
\end{eqnarray}
  is an  $\epsilon$-net in $M$.\par
\noindent  {\it(ii)} the open balls $B_{M}(p_{i},{\epsilon\over 2})$,
$i=1,\ldots,N$, are
disjoint, {\it i.e.}, $\{p_1,\ldots,p_N\}$
is a {\it minimal} $\epsilon$-net in $M$.\par
\end{Gaiad}
Similarly, upon
considering   the higher  order intersection patterns of the set
of balls $\{B_{M}(p_{i},\epsilon)\}$, we can define
the  two-skeleton ${\Gamma}^{(2)}(M)$, and eventually  the
nerve ${\cal N}\{B_i\}$ of the geodesic balls covering of   the
manifold   $M$:
\begin{Gaiad}
Let  $\{B_i(\epsilon)\}$ denote a minimal $\epsilon$-net in $M$.
The geodesic ball nerve ${\cal N}\{B_i\}$ associated with
$\{B_i(\epsilon)\}$ is the polytope whose
$k$-simplexes $p_{i_1i_2 \ldots i_{k+1}}^{(k)}$, $k=0,1,\ldots$, are
defined
by the collections of $k+1$ geodesic balls such that
$B_1 \cap {B}_2 \cap \ldots \cap {B}_{k+1} \not= \emptyset$. \par
\end{Gaiad}
Thus, for instance, the vertices $p_i^{(0)}$ of ${\cal N}\{B_i\}$
correspond
to
the balls $B_i(\epsilon)$; the edges $p_{ij}^{(1)}$ correspond to
pairs of
geodesic balls $\{B_i(\epsilon),B_j(\epsilon)\}$ having a not-empty
intersection $B_i(\epsilon) \cap B_j(\epsilon) \not= \emptyset$;
and the faces $p_{ijk}^{(2)}$ correspond to triples of
geodesic balls with not-empty intersection $B_i(\epsilon)\cap
{B}_j(\epsilon)\cap {B}_k(\epsilon) \not= \emptyset$. \par

\newtheorem{Gaiar}{Remark}
\begin{Gaiar}
 Notice that, in general, this polytope has a
dimension which is greater than the dimension $n$ of the underlying
manifold. However, as $\epsilon\to 0$, such dimension cannot grow
arbitrarily large being bounded above by a constant depending only on
$r$, $n$, and $D$, (see below).\par
\end{Gaiar}

Minimal geodesic ball coverings provide  a means for introducing a
short-distance cutoff as for a dynamical triangulation, while
hopefully
mantaining a more direct connection with the geometry and in
particular with the topology of the underlying manifold.
The basic observation here is that such coverings are
naturally labelled, (or coloured), by the fundamental groups of the
balls. Indeed,
according to the properties of the distance function, (see, for
instance [Ch]),
given ${\epsilon}_1<{\epsilon}_2\leq\infty$, if in
${\bar B_i({\epsilon}_2)}\backslash B_i({\epsilon}_1)$ there are no
critical points of the distance function ${\sigma}_i$, then this
region is homeomorphic to
${\partial}B_i({\epsilon}_1)\times[{\epsilon}_1,{\epsilon}_2]$,
and  ${\partial}B_i({\epsilon}_1)$ is a topological submanifold
without boundary.
One defines a {\it criticality radius},
${\epsilon}_i$, for each ball $B_i(\epsilon)$, as the largest
$\epsilon$ such that $B_i(\epsilon)$ is free of critical points.
Corresponding to such value of the radius $\epsilon$, the ball
$B_i({\epsilon})$ is
homeomorphic to an arbitrarily small open ball with center $p_i$,
and thus it is homeomorphic to a standard open ball.
It can be easily checked, through direct examples, that the
criticality radius of geodesic balls of manifolds in $\Ricco$ can
be arbitrarily small, (think of the geodesic balls drawn near the
rounded off tip of a cone), thus arbitrarily small metric balls in
manifolds of bounded geometry are not necessarily contractible, and
therefore, in general, the $B_i(\epsilon)$ are not homeomorphic to a
standard open ball.\par
\vskip 0.5 cm

\subsection{Connections with dynamical triangulations}

Since the geodesic ball coverings are to play an important role in our
development, a few remarks about
the connection between such coverings and dynamical triangulations are
in order.\par
\vskip 0.5 cm
As recalled in the introductory remarks, a dynamical triangulation of
a (pseudo)-manifold
can be used to produce a metric on that manifold, by declaring all the
simplexes in the triangulation isometric to the standard simplex of
the appropriate dimension, and by assuming that the edge lengths are
all equal to some fundamental length. An $n$-dimensional dynamical
triangulation is actually constructed by successively gluing pairs of
such flat $n$-simplexes along some of their
$(n-1)$-faces, until one gets a complex without boundary.
This gives a collection of compatible
metrics, on pieces of the resulting pseudo-manifold, which can be
extended to a  genuine
metric, since between any two points there is a path minimizing the
distance, (one speaks of a pseudo-manifolds since, for $n>2$, the
complex constructed by this gluing procedure may have some vertices
whose neighborhood is not homeomorphic to the standard euclidean
ball).\par
\vskip 0.5 cm
We identify two dynamical triangulations of the same underlying
manifold $M$ if there is a one-to-one mapping of vertices, edges,
faces, and higher dimensional simplexes of one onto vertices, edges,
faces, and higher dimensional simplexes of the other which preserves
incidence relations. If no such mapping exists the the two dynamical
triangulations are said to be {\it distinct}. Notice that sometimes
one says that such dynamical triangulations are combinatorially
distinct. Since this may be source of confusion, (in dynamical
triangulation theory the notion of combinatorial equivalence is
synonimus of PL-equivalence, see below), we carefully avoid the use of
the qualifier ``combinatorial'' in this context.\par
\vskip 0.5 cm

\noindent On a dynamical triangulation so constructed, one can define
metric balls, and consider minimal
geodesic ball coverings. Actually, it is clear that in a generic
metric space there are many
distinct
ways of introducing
minimal geodesic ball coverings with a given  radius of the balls. As
a simple example, consider a portion of Euclidean three-space, (one
may wish to identify boundaries so to obtain a
flat three-torus). It is well
known that a portion of Euclidean three-space can be
packed and covered, with small spheres of a given radius,
in many inequivalent ways, to the effect that in the limit, for ${\bf
R}^3$, there are
uncountably many such coverings.\par
\vskip 0.5 cm
As for dynamical triangulations, we
identify two geodesic ball coverings, $\{B_i\}_1$ and
$\{B_k\}_2$,
of the same underlying manifold $M$ if there is a one-to-one mapping
of vertices, edges, faces, and higher dimensional simplexes of the
nerve of $\{B_i\}_1$ onto vertices, edges, faces, and higher
dimensional simplexes of the nerve of
$\{B_k\}_2$,   which preserves incidence relations. If no such mapping
exists the the two geodesic ball coverings are said to be {\it
distinct}. \par
\vskip 0.5 cm

Generally, given a manifold triangulated with
$n$-dimensional simplexes with a given edge length,
 we can always introduce a minimal geodesic ball covering whose
properties are closely connected with the properties  of the
underlying
triangulation. This can be done  according to the following

\begin{Gaiad}
Let $(M,T)$ denote a manifold (a compact polyhedron) triangulated with
fixed edge
legth equilateral  simplexes, and let $\epsilon$ denote the length of
the edges. With each vertex $p_i$ belonging to the triangulation,  we
associate
the largest  open  metric ball contained in the open star of
$p_i$. Then the metric ball covering of $(M,T)$ generated by such
balls
$\{B_i\}$ is a minimal geodesic
ball covering. It defines
the geodesic ball covering associated with
the dynamically triangulated manifold $(M,T)$.
\end{Gaiad}
\vskip 0.5 cm

It is immediate to see that the set of balls considered defines
indeed a minimal geodesic ball covering. The open balls obtained from
$\{B_i\}$ by halfing their radius
are disjoint  being contained in the open stars
of $\{p_i\}$ in the baricentric subdivision of the triangulation.
The balls
with doubled
radius cover $(M,T)$, since they are the largest open balls
contained in the stars of the vertices $\{p_i\}$  of $T$.\par
\vskip 0.5 cm
In order to connect the enumeration of {\it distinct} geodesic ball
coverings with the enumeration of {\it distinct} triangulations, we
recall that any two dynamical triangulations are said to be {\it
Combinatorially Equivalent} if the two triangulations can be
subdivided into the same finer triangulation. In other words, if they
correspond to triangulations $T_1$ and $T_2$ of the same abstract
compact polyhedron $P$. This last remark follows since any two
triangulations of a compact polyhedron have a common subdivision.
Notice that quite often, when considering a particular triangulation
$(M,T)$ is standard usage to identify the abstract polyhedron $M$ with
$|T|$, the union of the cells of $T$, (the underlying polyhedron
associated with $T$). The more so when dealing with dynamical
triangulations, where the emphasis is on the actual construction of
$T$. This identification is a source of confusion in enumerative
problems and we shall keep distinct the abstract polyhedron $M$ from
$|T|$.
\vskip 0.5 cm
The relation between a dynamically triangulated manifold an the
associated geodesic ball covering implies the following

\vskip 0.5 cm
\newtheorem{Gaial}{Lemma}
\begin{Gaial}
If $(M,T_1)$ and $(M,T_2)$ are  any two distinct combinatorially
equivalent
fixed edge-length triangulations, then the corresponding geodesic ball
coverings $\{B_i\}_1$ and
$\{B_i\}_2$ are distinct.
\end{Gaial}
\vskip 0.5 cm
\noindent{\it Proof}.
This amounts to prove that
 the nerve associated with
 geodesic ball covering corresponding to a fixed edge-length
triangulation
is isomorphic, as a simplicial complex, to the given triangulation.
If this were not the case,
then, there should  be  at least one $k$-simplex in the nerve,
$p_{i_1\ldots i_{k+1}}^{k}$,
associated with the
mutual intersections of $k+1$ balls, for some $k>1$ , such that the
vertices of such $k$-simplex correspond to vertices of the
triangulation not connected by links.   But then
the corresponding balls ,
$B_i$, cannot mutually intersect, since they are contained in the
disjoint open stars of the respective vertices. Thus,
there cannot be such a simplex $p_{i_1\ldots i_{k+1}}^{k}$
to begin with.\par
\vskip 0.5 cm

In general, by choosing a different prescription for geodesic ball
coverings associated with fixed edge-length triangulations,
({\it e.g.}, by choosing differently the centers of the
balls $B_i$), we get a
nerve which is not necessarily isomorphic to
the dynamical triangulation itself. And, as already stressed, the
dimension
of the nerve is, in general, larger than the dimension
of the underlying manifold,
and even if we restrict our attention, say, to the four-skeleton, we
get a complex which is not the triangulation of a four-manifold.
 \vskip 0.5 cm
If we combine this remark with lemma 1 then we get the following
\newtheorem{Gaiap}{Proposition}
\begin{Gaiap}
For a given minimal short-distance cut-off, $\epsilon$,
the number of distinct geodesic ball
coverings is not smaller than the number of corresponding dynamical
triangulations.
\end{Gaiap}
\vskip 0.5 cm

Incidentally, by means of the above construction of a geodesic ball
covering associated with a given fixed edge-length triangulation, we
can also explain, in
terms of
dynamical triangulations, the origin of the possible non-trivial
topology of the balls.\par
Recall that in a $n$-dimensional simplicial manifold each vertex
has a sufficiently small neighborhood which is homeomorphic to the
standard $n$-dimensional euclidean ball. And, in such a case, the
above minimal
geodesic ball covering is necessarily
generated by contractible balls. Thus, non-contractible balls are
present if
we allow for dynamical triangulations associated with simplicial
pseudo-manifolds. And this is the typical case, at least in dimension
$n>2$, since
pseudo-manifolds are the natural outcome of the process of gluing
$n$-simplexes along their $(n-1)$-faces. \par
\vskip 0.5 cm
\subsection{Homotopy and geodesic ball coverings}
\vskip 0.5 cm
The above remarks suggest that one should be careful in understanding
in what sense, for $\epsilon$ sufficiently small, the geodesic ball
nerve gives rise to a polytope whose topology approximates the
topology of the manifold $M\in\Ricco$. This is a natural consequence
of the fact that the criticality radius for the geodesic balls is not
bounded below. In full generality, the geodesic ball nerve controls
only the homotopy type of the
manifold [GPW]. This follows by noticing that the inclusion of
sufficiently small geodesic balls into suitably larger balls is
homotopically trivial, and the geodesic ball nerve is thus a polytope
which is {\it homotopically dominating} the underlying manifold, {\it
viz.}, there exist maps
$f\colon{M}\to{\cal{N}}(B_i)$, and $g\colon{\cal{N}}(B_i)\to{M}$,
with $g\cdot{f}$ homotopic to the identity mapping in $M$.\par
It may appear rather surprising, but this homotopical control is more
than sufficient for yielding the entropic estimates we are looking
for.
\vskip 0.5 cm

On the geometrical side,  there are a wealth of good properties of
geodesic ball coverings which make them particularly appealing for
applications in simplicial quantum gravity. As a good start,
we can notice that the equivalence relation defined by manifolds
with
(combinatorially) isomorphic geodesic ball one-skeletons partitions
$\Ricco$  into disjoint equivalence classes whose
finite number can be estimated in terms of the parameters
$n$, $k$, $D$.  Each equivalence
  class of manifolds is
characterized by the abstract (unlabelled)    graph
${\Gamma}_{(\epsilon)}$
defined  by the  $1$-skeleton of the $L(\epsilon)$-covering. The
order of
any
 such graph ({\it i.e.}, the number of
vertices) defines  the {\it filling  function}
$N_{(\epsilon)}^{(0)}$,
while
the structure  of the  edge set  of ${\Gamma}_{(\epsilon)}$
defines  the (first order) intersection pattern $I_{(\epsilon)}(M)$ of
$(M,\{B_i(\epsilon)\})$.\par
\vskip 0.5 cm
It is important
to remark that  on $\Ricco$ neither the filling function nor the
intersection pattern can  be arbitrary. The filling function is
always
bounded
 above for each  given $\epsilon$, and  the best  filling,
with geodesic  balls  of  radius  $\epsilon$, of
a riemannian manifold of diameter $diam(M)$, and Ricci curvature
$Ric(M)\geq (n-1)H$, is controlled by the
corresponding filling of the geodesic ball of radius
$diam(M)$ on the  space form  of  constant  curvature given by $H$,
the bound being of the form [Gr1] $N^{(0)}_{\epsilon}\leq
N(n,H(diam(M))^2,(diam(M))/\epsilon)$.\par
\vskip 0.5 cm
The multiplicity of the first
intersection pattern is  similarly  controlled  through  the
geometry
 of the  manifold to  the effect  that the  average degree,
$d(\Gamma)$,  of  the  graph ${\Gamma}_{(\epsilon)}$, ({\it
i.e.}, the  average number of edges incident on a
vertex of  the graph), is bounded above by  a constant  as the
radius of the balls defining  the covering  tend to  zero, ({\it
i.e.},
as $\epsilon \to 0$ ).  Such constant  is independent  from
$\epsilon$
and can be estimated [Gp] in terms of the parameters $n$, and
$H(diam(M))^2$, (it is this boundedness of the order of the geodesic
ball coverings that allows for the control of the dimension of the
geodesic ball nerve).\par
\vskip 0.5 cm
As expected, the filling function can be also related to the volume
$v=Vol(M)$
of the underlying manifold $M$. This follows by noticing
that [Zh]
for any manifold $M\in\Ricco$ there exist constants $C_1$ and $C_2$,
depending only on $n$, $r$, $D$, $V$, such that, for any $p\in M$, we
have
\begin{eqnarray}
C_1{\epsilon}^n\leq Vol(B_{\epsilon}(p))\leq C_2{\epsilon}^n
\end{eqnarray}
with $0\leq \epsilon \leq D$, (actually, here and in the previous
statements a lower bound on the
Ricci curvature
suffices). Explicitly, the constants $C_1$ and $C_2$ are provided by
\begin{eqnarray}
C_1\equiv \frac{V}{Vol^r(B(D))}{\inf}_{0\geq{\epsilon}\geq{D}}
\frac{1}{{\epsilon}^n}\int_0^{\epsilon}(\frac{\sinh\sqrt{-rt}}{\sqrt{-
r}})^{n-1}dt
\end{eqnarray}
and
\begin{eqnarray}
C_2\equiv {\sup}_{0\geq{\epsilon}\geq{D}}
\frac{1}{{\epsilon}^n}\int_0^{\epsilon}(\frac{\sinh\sqrt{-rt}}{\sqrt{-
r}})^{n-1}dt
\end{eqnarray}
where $Vol^r(B(D))$ denotes the volume of the geodesic ball of
radius D in the (simply connected) space of constant curvature $-r$,
and $D$, $r$, $V$, $n$ are the parameters characterizing the space of
bounded geometries $\Ricco$ under consideration.\par

Thus, if $v$ is the given volume of the underlying manifold $M$, by
the Bishop-Gromov relative comparison volume theorem we obtain that
there
exists a function ${\rho}_1(M)$, depending on $n$, $r$, $D$, $V$, and
on the actual geometry of the manifold $M$, with
$C_1\leq ({\rho}_1(M))^{-1}\leq C_2$, and such that, for any $m\geq
m_0$, we can write
\begin{eqnarray}
N^{(0)}_{\epsilon}(M) =v{\rho}_1(M){\epsilon}^{-n}
\label{volumemme1}
\end{eqnarray}
\vskip 0.5 cm

We conclude this section by recalling the following basic
finiteness results. They provide the
topological rationale underlying the use of
spaces
of bounded geometries in simplicial quantum gravity. We start with a
result expressing finiteness of homotopy types of manifolds of
bounded geometry [GPW]

\newtheorem{Gaiat}{Theorem}
\begin{Gaiat}
For any dimension $n \geq 2$, and  for $\epsilon$
sufficiently small, manifolds in $\Ricco$ with the same
geodesic ball
$1$-skeleton ${\Gamma}_{(\epsilon)}$ are homotopically equivalent, and
the number of different homotopy-types of manifolds
realized in $\Ricco$ is finite and is a function of $n$,
$V^{-1}D^n$, and $rD^2$.
\end{Gaiat}
(Two manifolds $M_1$ and $M_2$ are said to have the same
 homotopy type if there exists a continuous
 map $\phi$ of $M_1$ into $M_2$ and $f$ of $M_2$ into $M_1$,
such that both $f \cdot \phi$ and $\phi \cdot f$ are homotopic
to the respective identity mappings, $I_{M_1}$ and $I_{M_2}$.
 Obviously, two homeomorphic manifolds are of
the same homotopy type, but the converse is not true).\par
\vskip 0.5 cm
Notice that in dimension three one can replace the lower bound of the
sectional curvatures with a lower bound on the Ricci
curvature [Zh]. Actually, a more general topological finiteness
theorem can be stated under a rather weak condition of local geometric
contractibility. Recall that a  continuous function
${\psi}\colon [0,\alpha)\to {\bf R}^+$, $\alpha >0$, with
${\psi}(0)=0$, and ${\psi}(\epsilon)\geq \epsilon$, for all
$\epsilon\in [0,\alpha)$, is a local geometric contractibility
function for a riemannian manifold $M$ if, for each $x\in M$ and
$\epsilon\in (0,\alpha)$, the open ball $B(x,\epsilon)$ is
contractible in $B(x,{\psi}(\epsilon))$ [GrP],
(which says that a small ball is contractible relative to a bigger
ball). Given a local geometric contractibility function  one obtains
the
following [GrP]
\begin{Gaiat}
Let ${\psi}\colon [0,\alpha)\to {\bf R}^+$, $\alpha >0$, be a
continuous
function with ${\psi}(\epsilon)\geq \epsilon$ for all $\epsilon\in
[0,\alpha)$ and such that, for some constants $C$ and $k\in (0,1]$, we
have the growth condition ${\psi}(\epsilon)\leq C{\epsilon}^k$, for
all $\epsilon\in [0,\alpha)$. Then for each $V_0>0$ and $n\in {\bf
R}^+$ the class ${\cal C}(\psi, V_0,n)$ of all compact $n$-dimensional
Riemannian manifolds with volume $\leq V_0$ and with $\psi$ as a local
geometric contractibility function contains\par
\noindent {\it (i)} finitely many simple homotopy types (all $n$),\par
\noindent {\it (ii)} finitely many homeomorphism types if $n=4$,\par
\noindent {\it (iii)} finitely many diffeomorphism types if $n=2$ or
$n\geq 5$.
\end{Gaiat}
Actually the growth condition on $\psi$ is necessary in order to
control the dimension of the limit spaces resulting from
Gromov-Hausdorff convergence of a sequence of manifolds in
${\cal C}(\psi, V_0,n)$. As far as homeomorphism types are concerned,
this condition can be removed [Fe]. Note moreover that
infinite-dimensional limit spaces cannot occur in the presence of a
lower bound on sectional curvature as for manifolds in $\Ricco$.
Finiteness of the homeomorphism types cannot be proved in dimension
$n=3$ as long as the Poincar\'e conjecture is not proved. If there were
a fake three-sphere then one could prove [Fe] that a statement
such as {\it (ii)} above is false for $n=3$.
Finally, the statement on finiteness of {\it simple homotopy} types,
in
any dimension, is particularly important for the applications in
quantum gravity we discuss in the sequel. Roughly speaking the notion
of simple
homotopy is a refinement of the notion of homotopy equivalence, and it
may
be thought of as an intermediate step between homotopy equivalence and
homeomorphism.
\vskip 0.5 cm

The machinery needed to characterize the entropy function for geodesic
ball coverings in four-dimensional manifolds of bounded geometry is
now at hand.

\section{Topology and Entropy of metric ball coverings}

The combinatorial structure associated with minimal geodesic ball
coverings appears more complex than the combinatorial structure of
dynamical triangulations. However, the counting of all possible
distinct minimal geodesic ball coverings of given radius, on a
manifold of bounded geometry, is more accessible than the counting of
distinct dynamical triangulations.\par
This fortunate situation arises because we can label the balls
$B_{\epsilon}(p_i)$
 with their non-trivial fundamental group
${\pi}_1(B_{\epsilon}(p_i);p_i)$, (obviously, since we are interested
to the distinct classes of covering, we need to
factor out the trivial labelling associated with the centers $p_i$ of
the balls). Thus the counting problem we face is basically the
enumeration of such  inequivalent
topological labellings of the balls of the covering. Such an
enumeration is not yet very accessible.
As it stands, there are constraints on
the fundamental groups
${\pi}_1(B_{\epsilon}(p_i);p_i)$, expressed by Seifert-VanKampen's
theorem, which express the match between the intersection pattern of
the balls and the homomorphisms
${\pi}_1(\cap_iB_{\epsilon}(p_i);x_0)\to{\pi}_1(M;x_0)$ associated
with the injection of clusters of mutually intersecting balls into
$M$, ($x_0$ being a base point in
$M\cap_iB_{\epsilon}(p_i)$).
Such difficulties can be circumvented by using as labels, rather than
the fundamental groups
themselves, a
cohomology with local coefficients in representations of
${\pi}_1(B_{\epsilon}(p_i);p_i)$ into a Lie group $G$.
Roughly speaking, this means that we are using {\it flat bundles
corresponding to the representation $\theta$} as non-trivial
labels for the balls.\par
This construction gives to the counting problem of inequivalent
geodesic ball coverings an unexpected interdisciplinary flavour which
blends in a nice way riemannian geometry ({\it the metric properties
of the balls}), topology ({\it the action of the fundamental group on
homology}), and representation theory ({\it the structure of the space
of inequivalent representation of the fundamental
group of a $n$-dimensional manifold, $n\geq{3}$}), into a Lie group $G$.
\vskip 0.5 cm
We wish to stress that  a
similar approach may be suitable also for a direct enumeration
of  dynamical triangulations
since  flat bundles on the simplexes, (again associated with
representations of the fundamental group of the underlying
PL-manifold), do provide a
 natural topological labelling of the simplexes. It is true that
such simplexes have no non-trivial topology, (they are contractible),
and that a flat bundle, (associated with the given representation
$\theta$), over one simplex is isomorphic to the flat bundle
over any other simplex. However, such isomorphism is not canonical,
as is obviously shown by the fact that one may get a non-trivial
flat bundle by gluing such local bundles if the underlying manifold
has a non-trivial fundamental group, (we wish to thank J.
Ambj\o rn and B. Durhuus for discussions that draw our attention
to this further possibility).

\vskip 0.5 cm
\subsection{Cohomology with local coefficients and representation
spaces}

In order to describe either the topological aspects or the basic
properties of the representation spaces mentioned above and  which
play a prominent role into our entropy estimate, it will be convenient
to recall some basic facts about cohomology with local
coefficients. Details of the theory summarized here can be found in
[DNF],[RS], [JW].\par
\vskip 0.5 cm
Let $(M,\{B_{\epsilon}(p_i)\})\in\Ricco$ be a manifold of bounded
geometry endowed with a minimal geodesic ball covering, and thought of
as a cellular or simplicial complex, (for instance by associating with
$(M,\{B_{\epsilon}(p_i)\})$ the corresponding nerve
${\cal N}$; in what follows we tacitly exploit the fact that a
sufficiently fine minimal geodesic ball covering has the same homotopy
type of the underlying manifold).
We let ${\pi}_1(M)$ denote the
fundamental group of $(M,\{B_{\epsilon}(p_i)\})$.
Such ${\pi}_1(M)$ is finitely
generated, and can be assumed to be finitely presented.\par
\vskip 0.5 cm
Let $\hat{M}\to{M}$ denote the universal covering of $M$,
on which ${\pi}_1(M)$ acts by deck transformations.
Let us introduce the homology complex
$C_*(\hat{M})=\bigoplus_{i\in{\Bbb N}}C_i(\hat{M})$ where the
 chains in
$C_i(\hat{M})$ are of the form
$\sum_{j,\gamma}{\lambda}_{j\gamma}a_j(\hat{\sigma}^i_{\gamma})$
where ${\lambda}_{j\gamma}$ are integers, $a\in {\pi}_1(M)$,
and $\hat{\sigma}^i_{\gamma}$ are a set of chosen $i$-cells in
$\hat{M}$. This is tantamount to say that the chains
$C_i(\hat{M})$ have coefficient in the group ring
${\bf Z}{\pi}_1(M)$, {\it i.e.}, in the set of all finite
formal sums $\sum{n}_ia_i$, $n_i\in{\bf Z}$,
$a_i\in{\pi}_1(M)$, with the natural definition of addition and
multiplication.
\vskip 0.5 cm
Let ${\theta}\colon{\pi}_1(M)\to{G}$, be a representation of
${\pi}_1(M)$ in a Lie group $G$ whose Lie
algebra ${\frak g}$ carries an
$Ad$-invariant, symmetric, non-degenerate bilinear form, ({\it i.e.},
a metric).
The representation $\theta$ defines a
flat bundle, that we denote by ${\frak g}_{\theta}$.
This bundle is constructed by
exploiting  the adjoint representation of $G$ on its
Lie algebra ${\frak g}$, {\it i.e.},
$Ad\colon{G}\to{End({\frak g})}$, and by considering the action of
${\pi}_1(M)$ on ${\frak g}$ generated by composing the adjoint action
and the representation $\theta$:
\begin{eqnarray}
{\frak g}_{\theta}=\hat{M}\times{\frak
g}/{\pi}_1\otimes[Ad({\theta}(\cdot))]^{-1}
\end{eqnarray}
where ${\pi}_1\otimes[Ad({\theta}(\cdot))]^{-1}$ acts, through
${\pi}_1(M)$, by deck transformations on $\hat{M}$ and by
$[Ad({\theta}(\cdot))]^{-1}$ on the Lie algebra ${\frak g}$. More
explicitly, if $\hat{\sigma}_1$, $\hat{\sigma}_2$ are cells in
$\hat{M}$, and $g_1$, $g_2$ are elements of ${\frak g}$, then
\begin{eqnarray}
(\hat{\sigma}_1,g_1)\sim (\hat{\sigma}_2,g_2)
\end{eqnarray}
if and only if
\begin{eqnarray}
\hat{\sigma}_2=\hat{\sigma}_1a
\end{eqnarray}
and
\begin{eqnarray}
g_2=[Ad({\theta}(a))]^{-1}g_1
\end{eqnarray}
for some $a\in{\pi}_1(M)$.\par
\vskip 0.5 cm
In this way we can define a cellular chain complex
$C_*(M,{\frak g}_{\theta})$ with  coefficients in the flat bundle
${\frak g}_{\theta}$. First we consider chains with coefficients in
the
Lie algebra ${\frak g}$, {\it viz.},
\begin{eqnarray}
\sum_jg_j\hat{\sigma}^i_j
\end{eqnarray}
with $g_j\in{\frak g}$, and then quotient the resulting chain
complex $C(\hat{M})\otimes{\frak g}$ by the action of
${\pi}_1\otimes[Ad({\theta}(\cdot))]^{-1}$. This yields
for an action of ${\pi}_1(M)$ on the above chains expressed by
\begin{eqnarray}
a(\sum_jg_j\hat{\sigma}^i_j)\to
\sum_j([Ad({\theta}(a))]^{-1}g_j)a(\hat{\sigma}^i_j)
\end{eqnarray}
for any $a\in{\pi}_1(M)$, ({\it i.e.}, we are considering
${\frak g}$ as a ${\pi}_1(M)$-module).
This action commutes with the boundary operator, and
as a consequence of the definition of the flat bundle
${\frak g}_{\theta}$,
the resulting homology $H_*(M,{\frak g}_{\theta})$ can be thought of
as
a homology
with local coefficients  in the flat bundle
${\frak g}_{\theta}$. By dualizing one defines the
cohomology $H^*(M,{\frak g}_{\theta})$, which enjoys the usual
properties of a cohomology theory.
Sometimes, for ease of
notation, we shall denote $H_*(M,{\frak g}_{\theta})$  and
$H^*(M,{\frak g}_{\theta})$ by $H^{{\frak g}}_*(M)$ and
$H_{{\frak g}}^*(M)$, respectively.\par
\vskip 0.5 cm
Let $B_{\epsilon}(p_h)$ be the generic ball of the covering
$(M,\{B_{\epsilon}(p_i)\})$. If we denote by
${\phi}_h\colon{\pi}_1(B_{\epsilon}(p_h);p_h)\to{\pi}_1(M;p_h)$
the homomorphism induced by the obvious inclusion map, then together
with $\theta$, we may also consider the representations
\begin{eqnarray}
{\theta}_h\colon
{\pi}_1(B_{\epsilon}(p_h);p_h)\to{\pi}_1(M;p_h)
\to{G}
\end{eqnarray}
obtained by composing $\theta$ with the various homomorphisms
${\phi}_h$ associated with the balls of the covering.\par
Notice that since arbitrarily small metric balls in manifolds
$M\in\Ricco$
can be topologically rather complicated, it cannot be excluded a
priori that the image

${\phi}_h[{\pi}_1(B_{\epsilon}(p_h);p_h)]$ in ${\pi}_1(M)$, (or more
generally in the fundamental group of a larger, concentric ball), has
an infinite number of  generators. However, as follows from a result
of S-h.Zhu,  in order to avoid such troubles
it is sufficient to choose the radius of the balls small enough [Zh]
\begin{Gaiat}
There are constants $R_0$, ${\epsilon}_0$ and $C$ depending only
on $n$, $r$, $D$, $V$, such that for any manifold $M\in\Ricco$,
$p\in{M}$, $\epsilon\geq{\epsilon}_0$, if
$i\colon{B_{\epsilon}(p)}\to{B_{R_0\epsilon}(p)}$ is the inclusion,
then any subgroup $K$ of
${\phi}_i({\pi}_1(B_{\epsilon}(p)))$ satisfies
$order(K)\leq{C}$.
\end{Gaiat}
Thus in particular, there is no element of infinite order in
${\phi}_i({\pi}_1(B_{\epsilon}(p)))$ whenever
$\epsilon\geq{\epsilon}_0$.\par
\vskip 0.5 cm

According to this latter result, by choosing
$\epsilon\geq{\epsilon}_0$ and by using the representations
${\theta}_h$, we may define the cohomologies
$H^*_{{\frak g}}(B_{\epsilon}(p_h))$ with local coefficients in the
corresponding flat bundles
${\frak g}_{\theta}|(B_{\epsilon}(p_h))$
defined
over the balls $B_{\epsilon}(p_h)$.
 As labels, these cohomology groups
are easier to handle than the fundamental groups
${\pi}_1(B_{\epsilon}(p_i))$. This is so because the constraints we
have to implement on the intersections of the balls,
relating $\{H^*_{{\frak g}}(B_{\epsilon}(p_i))\}_i$ to
$H^*_{{\frak g}}(M)$, are simply obtained by iterating the cohomology
long exact Mayer-Vietoris sequence obtained from the short exact
sequences connecting
the cochains $C^*_{{\frak g}}(B_{\epsilon}(p_i))$,
$C^*_{{\frak g}}(\cup{ B_{\epsilon}(p_i}))$, and
$C^*_{{\frak g}}(\cap{B_{\epsilon}(p_i)})$. For instance, given any
two
intersecting balls $B(p_i)$ and $B(p_h)$, we get
\begin{eqnarray}
0\to C^j_{{\frak g}}(B(p_i)\cup B(p_h))\to
C^j_{\frak{g}}(B(p_i))\oplus
C^j_{\frak{g}}(B(p_h))\to
C^j_{\frak{g}}(B(p_i)\cap B(p_h))\to 0
\end{eqnarray}
whose corresponding cohomology long exact sequence reads
\begin{eqnarray}
\ldots\to H^j_{\frak{g}}(B(p_i)\cup B(p_h))&\to
H^j_{\frak{g}}(B(p_i))\oplus
H^j_{\frak{g}}(B(p_h))\to\nonumber\\
\to H^j_{\frak{g}}(B(p_i)\cap B(p_h))&\to
H^{j+1}_{\frak{g}}(B(p_i)\cup
B(p_h))\to\ldots
\end{eqnarray}
Similar expressions can be worked out for any cluster
$\{B_{\epsilon}(p_i)\}_{i=1,2,\ldots}$ of intersecting geodesic balls,
(see section 3.3),
and they can be put at work for our counting purposes
by introducing the Reidemeister torsion,
a graded version of the absolute value of the
determinant of  an isomorphism of vector spaces.
\vskip 0.5 cm
\subsection{Torsions}

Let us start by recalling that
 by hypothesis $\frak{g}$ is endowed with an Ad-invariant,
symmetric, non-degenerate bilinear form, ({\it i.e.}, with a
metric), thus we can introduce  orthonormal bases,
$\{X_k\}_{k=1,\ldots,dim(G)}$, for the Lie algebra
$\frak{g}$.
Since the adjoint representation is an
orthogonal representation of $G$ on $\frak{g}$,  we can introduce a
volume element on the cochain complex
$C^*(M,\frak{g}_{\theta})$, by exploiting such orthonormal bases: by
identifying $C^i(M,\frak{g}_{\theta})$
with a direct sum of a copy of $\frak{g}$ for each $i$-cell
$\hat{\sigma}^i_j$ in $\hat{M}$, we can take
$\hat{\sigma}^i_J\otimes{X_k}$
as an orthonormal basis of $C^i(M,\frak{g}_{\theta})$ and define
the space of volume forms as the determinant line
$detline|C^*(M,{\frak g}_{\theta})|\equiv
\prod_{i}(detline|C^i(M,{\frak g}_{\theta})|)^{(-1)^i}$, where
$detline|C^i(M,{\frak g}_{\theta})|$ denotes the  line  of volume
elements on
$C^i(M,{\frak g}_{\theta})$ generated by all possible choices of the
orthonormal bases $\hat{\sigma}^i_J\otimes{X_k}$.
 Explicitly, if on
each $C^i(M,{\frak g}_{\theta})$ we chose the volume forms
$t_i$, then the corresponding volume element is obtained by setting
\begin{eqnarray}
t({\frak g}_{\theta})=\prod_i(t_i)^{(-1)^i}\in
detline|C^*(M,{\frak g}_{\theta})|
\end{eqnarray}
\vskip 0.5 cm
Let $d_i\colon{C_{i}^{\frak g}}\to{C_{i+1}^{\frak g}}$ be the
coboundary
operator in $C^*_{\frak g}$, and as usual let us denote by
$Z_{\frak g}^i\equiv ker(d)$,
$B^i_{\frak g}\equiv Im(d)$. From the
short exact sequences
\begin{eqnarray}
0\to{Z^{i}_{\frak{g}}}\to{C^{i}_{\frak{g}}}\to{B^{i+1}_{\frak{g}}}
\to{0}
\end{eqnarray}
and
\begin{eqnarray}
0\to{B^{i}_{\frak{g}}}\to{Z^{i}_{\frak{g}}}\to{H^{i}_{\frak{g}}}
\to{0}
\end{eqnarray}
we respectively get that there are  natural isomorphisms
\begin{eqnarray}
{\Lambda}^{dim(Z^i)}{Z^{i}_{\frak{g}}}\otimes
{\Lambda}^{dim(B^{i+1})}{B^{i+1}_{\frak{g}}}\to
{\Lambda}^{dim(C^i)}{C^{i}_{\frak{g}}}
\end{eqnarray}
and
\begin{eqnarray}
{\Lambda}^{dim(B^i)}{B^{i}_{\frak{g}}}\otimes
{\Lambda}^{dim(H^{i})}{H^{i}_{\frak{g}}}\to
{\Lambda}^{dim(Z^i)}{Z^{i}_{\frak{g}}}
\end{eqnarray}
where ${\Lambda}^{dim(\cdot)}$ denotes the top dimensional exterior
power on the vector space considered, (recall that if
$\ldots\to{V_i}\to{V_{i+1}}\to\ldots$ is a finite exact sequence, then
there is a  canonical isomorphism
$\otimes_{i-even}{\Lambda}^{dim(V_i)}V_i=
\otimes_{i-odd}{\Lambda}^{dim(V_i)}V_i$).
It follows that there is an
isomorphism
\begin{eqnarray}
{\Lambda}^{dim(B^i)}{B^{i}_{\frak{g}}}\otimes
{\Lambda}^{dim(H^{i})}{H^{i}_{\frak{g}}}
\otimes{\Lambda}^{dim(B^{i+1})}{B^{i+1}_{\frak{g}}}\to
{\Lambda}^{dim(C^i)}{C^{i}_{\frak{g}}}
\end{eqnarray}
This isomorphism is explicitly realized by
fixing an orthonormal bases ${\bf h}^{(i)}$, and
${\bf b}^{(i)}$ for
${H^{i}_{\frak{g}}}$ and ${B^{i}_{\frak{g}}}$, respectively.
Thus, if we denote by
${\nu}_i\equiv{\wedge}_q^{dim(H)^i}h^{(i)}_q$
the corresponding volume form in
${H^{i}_{\frak{g}}}$ (lifted to ${C^{i}_{\frak{g}}}$),
we can write
\begin{eqnarray}
[{\wedge}_q^{dim(B)^i}b^{(i)}_q]\wedge
[{\wedge}_q^{dim(B)^{i+1}}{d}b^{(i+1)}_q]\wedge
[{\wedge}_q^{dim(H)^i}h^{(i)}_q]=t_i({\nu}_i)
[{\wedge}^{dim(C)^i}\hat{\sigma}^i_j\otimes{X_k}]
\end{eqnarray}
for some scalar $t_i({\nu}_i)\not= 0$.
\vskip 0.5 cm
With these  remarks out of the way,
and setting, for notational convenience,
${\mu}_i\equiv{\wedge}^{dim(C)^i}\hat{\sigma}^i_j\otimes{X_k}$,
we can define the
Reidemeister torsion associated with the cochain complex
$C^*_{\frak{g}}$ according to the following
\begin{Gaiad}
For a given choice of volume elements
${\nu}_i$ in ${H^*_{\frak{g}}}$,
the torsion of the cochain complex $C^*_{\frak{g}}$ is the volume
element
\begin{eqnarray}
{\Delta}^{\frak g}(M;{\bf{\mu}},{\bf{\nu}})\equiv
t({\frak g}_{\theta})=\prod_i[t_i({\nu}_i)]^{(-1)^i}\in
detline|C^*(M,{\frak g}_{\theta})|
\end{eqnarray}
\end{Gaiad}
\vskip 0.5 cm
Notice that we have  selected a particular definition
out of many naturally equivalent ones, (see [RS],
for a more detailed treatment).\par
\vskip 0.5 cm

As the notation suggests, it is easily checked that
${\Delta}^{\frak g}(M;{\bf{\mu}},{\bf{\nu}})$  is independent of the
particular choice of the bases $\{\bf b\}^{(i)}$ for the
${B^{i}_{\frak{g}}}$. Moreover, if the complex
${C^{i}_{\frak{g}}}$ is acyclic then
${\Delta}^{\frak g}(M;{\bf{\mu}},{\bf{\nu}})$ is also independent
from the choice of a volume element in ${H^*_{\frak{g}}}$,
(recall  that the cochain complex ${C^{i}_{\frak{g}}}$ is said to be
acyclic if ${H^{i}_{\frak{g}}}=0$ for all $i$).
\vskip 0.5cm
Obviously, we may have worked as well in homology $H_*^{\frak g}$, by
obtaining ${\Delta}^{\frak g}(M;{\bf{\mu}},{\bf{\nu}})$ as an
element of  $detline|C_*(M,{\frak g}_{\theta})|$ depending now on a
choice of volume elements ${\nu}^i$ in the homology
groups $H_i^{\frak g}$.\par
\vskip 0.5 cm
It is important to stress that if the complex
$C^*(M,{\frak g}_{\theta})$  is not acyclic then
${\Delta}^{\frak g}(M;{\bf{\mu}},{\bf{\nu}})$ is not a scalar but
a volume element in $detline|H^*(M,{\frak g}_{\theta})|$ under the
natural identification between this latter line bundle and
$detline|C^*(M,{\frak g}_{\theta})|$.\par

\vskip 0.5 cm
The torsion is an interesting combinatorial invariant of a complex,
since it is invariant under subdivision of $M$ and it is
deeply related to homotopy theory. In particular, given a homotopy
equivalence $f\colon(M_1,{\cal N}_1)\to(M_2,{\cal N}_2)$ between two
cellular complexes, there is a  correspondence between the flat
bundles over $M_1$ and the flat bundles over
$M_2$ induced by the isomorphism ${\pi}_1(M_1)\to{\pi}_1(M_2)$
and by the representation ${\theta}$ of such groups into the Lie
group $G$. However, the corresponding torsions are not necessarily
equal, this
being the case if and only if $h$ is (homotopic to) a Piecewise-Linear
(PL) equivalence between the complexes in question.\par
\vskip 0.5 cm
Also notice that if the manifold $M$ underlying the complex is an
orientable,
even dimensional manifold {\it without boundary} and the cochain
complex $C_*(M,{\frak g}_{\theta})$ is acyclic, then
${\Delta}^{\frak g}(M;{\bf{\mu}},{\bf{\nu}})=1$, (see {\it e.g.},
[Ch]). Thus it would seem that calling into play such invariant,
for counting geodesic ball coverings over a four dimensional manifold,
is useless. However, there are three reasons which show that the role
of torsion is not so trivial
for our counting purposes. First, we shall deal with the torsions of
the geodesic balls which are four dimensional manifolds with a
non-trivial three-dimensional boundary. Moreover,  the
complexes  we need to use are not acyclic, and the behavior of
volume elements ${\bf{\nu}}$ in cohomology will play a basic role.
Finally, the fact that we are in dimension  four will be imposed only
in the final part of our paper, when estimating the dimension of the
tangent space to the set of all conjugacy classes of representations
of the fundamental group, (see below). In this connection, we wish to
stress that {\it the analysis which follows holds for for any
$n$-dimensional manifold $M\in\Ricco$
with $n\geq 2$}.
\vskip 0.5 cm

We now examine the dependence of
${\Delta}^{\frak g}(M;{\bf{\mu}},{\bf{\nu}})$ on the particular
representation ${\theta}\colon{\pi}_1(M)\to{G}$. To this end,
let $\frac{Hom({\pi}_1(M),G)}{G}$ denote the set of all conjugacy
classes of representations of the fundamental group ${\pi}_1(M)$
into the Lie group $G$. Notice that   if $\theta$ and
$F{\theta}F^{-1}$ are two conjugate representations of ${\pi}_1(M)$ in
$G$, then
through the map $Ad(F)\colon{\frak g}\to{\frak g}$ we get a natural
isomorphism between the groups $H_{i}(M,{\frak g})$ and
$H_{i}(M,F{\frak g}F^{-1})$. Thus it follows that
the torsion corresponding to the representation $\theta$ and the
torsion corresponding to the conjugate representation
$F{\theta}F^{-1}$ are equal, and ${\Delta}^{\cal
G}(M;{\bf{\mu}},{\bf{\nu}})$ is actually well defined on the
conjugacy class of representations
$[\theta]\in\frac{Hom({\pi}_1(M),G)}{G}$.
\vskip 0.5 cm
When  defining the Reidemeister torsion, one of
the advantages of using the homology $H_*(M,{\frak g})$ with local
coefficients in the
bundle ${\frak g}_{\theta}$ lies in the fact that the corresponding
cohomology is strictly related to the structure
of the representation space $\frac{Hom({\pi}_1(M),G)}{G}$.
This point is quite important,
since we are interested in understanding the dependence
of ${\Delta}^{\frak g}(M;{\bf{\mu}},{\bf{\nu}})$
when deforming the particular
representation ${\theta}\colon{\pi}_1(M)\to{G}$ through a
differentiable one-parameter family of representations
${\theta}_t$ with ${\theta}_0=\theta$ which are not tangent to the
$G$-orbit of $\theta\in{Hom({\pi}_1(M),G)}$.\par
\vskip 0.5 cm
To this end,
let us rewrite, for $t$ near $0$, the given one-parameter family of
representations ${\theta}_t$ as [Go], [Wa]
\begin{eqnarray}
{\theta}_t=\exp[tu(a)+O(t^2)]{\theta}(a)
\end{eqnarray}
where $a\in {\pi}_1(M)$, and where $u\colon{\pi}_1(M)\to{\frak g}$. In
particular,
 given $a$ and $b$ in ${\pi}_1(M)$, if we differentiate the
homomorphism condition ${\theta}_t(ab)={\theta}_t(a){\theta}_t(b)$, we
get that
$u$ actually is a one-cocycle of ${\pi}_1(M)$ with coefficients
in the ${\pi}_1(M)$-module ${\frak g}_{\theta}$, {\it viz.},
\begin{eqnarray}
u(ab)=u(a)+ [Ad({\theta}(a))]u(b)
\end{eqnarray}
Moreover, any $u$ verifying the above cocycle condition leads to a map
${\theta}_t\colon{\pi}_1(M)\to{G}$ which, to first order in $t$,
satisfies the homomorphism condition. This remark implies that the
(Zariski) tangent space to $Hom{({\pi}_1(M),G)}$ at
$\theta$, can be identified with $Z^1(M,{\frak g}_{\theta})$.\par
In a similar way, it can be shown that the tangent space to the
$Ad$-orbit through ${\theta}$ is
$B^1(M,{\frak g}_{\theta})$. Thus, the (Zariski) tangent space to
$\frac{Hom({\pi}_1(M),G)}{G}$ corresponding to the conjugacy
class of representations
$[\theta]$ is $H^1(M,{\frak g}_{\theta})$.
And, as it is usual in deformation theory, this is the formal tangent
space to the representation space.\par
\vskip 0.5 cm
It must be emphasized that, in general, there are
obstructions [Go]
that do not allow the identification between the Zariski tangent
space with the actual tangent space to
$\frac{Hom({\pi}_1(M),G)}{G}$.
Typically we have troubles in correspondence to reducible
representations.  Since the tangent space to the isotropy
group of the representation $\theta$, is
$H^0(M,{\frak g}_{\theta})$, it follows that
$H^0(M,{\frak g}_{\theta})\not= 0$ precisely when there are reducible
representations. Further obstruction to identifying
$H^1(M,{\frak g}_{\theta})$ to the actual tangent space are in
$H^2(M,{\frak g}_{\theta})$. In deformation theory it is well known
that this space is to contain the obstructions to extend a formal
deformation to a finite deformation, ({\it i.e.}, in a language more
familiar to relativists, $H^2(M,{\frak g}_{\theta})$
is associated to a linearization instability around the given
representation in $\frac{Hom({\pi}_1(M),G)}{G}$). The triviality of
this space at a (conjugacy class of a) representation $\theta$ is a
necessary condition for ${\theta}$ to be a regular point of the
representation space $\frac{Hom({\pi}_1(M),G)}{G}$, and for
identifying
$H^1(M,{\frak g}_{\theta})$ with
$T_{\theta}[\frac{Hom({\pi}_1(M),G)}{G}]$.
\vskip 0.5 cm
We shall be ignoring the singularities produced by reducible
representations by restricting our considerations to the set of
irreducible  representations
${\cal S}\in\frac{Hom({\pi}_1(M),G)}{G}$, yet, in general we do not
assume that  $H^2(M,{\frak g}_{\theta})=0$. Not considering reducible
representations
is certainly not topologically justified in general, but is not yet
clear how to circumvent the difficulties associated with them.
Moreover, the results we obtain are well-defined in considerable
generality and do not seem to suffer too much by such restrictions.
\vskip 0.5 cm

Recall that $\Hom(\pi_1(X),G)^{\hbox{irr}}$ is a smooth analytic
submanifold of
$G^m$, for some $m$.  This provides ${\S}$ with an analytic structure,
possibly
outside some singular points.
Let $\S_0$ denote the smooth locus of ${\S}$, and let $d$ be its
dimension.

To be definite, we set $G=U(n)$. If we assume $M$ to be oriented,
the space
$\S_0$, regarded as the space of gauge equivalence classes of
flat connections $\nabla$ on  ${\frak g}_{\theta}$,  sits
inside both $\M_+$ and $\M_-$, these two spaces being the moduli
spaces of selfdual and anti-selfdual irreducible instantons on $M$.

Since
\begin{eqnarray}
\dim\M_-=\dim G(b_1-b_+-1)\,,\qquad
\dim\M_+=\dim G(b_1-b_--1)\,,
\end{eqnarray}
we get the inequalities
\begin{eqnarray}
d\leq \dim G(b_1-b_+-1),\qquad d\leq\dim G(b_1-b_--1)\,;
\label{uno}
\end{eqnarray}
by summing the two  inequalities we get
\begin{eqnarray}
d\leq -\unmezzo\dim G\chi(M)\,.
\end{eqnarray}
 We stress that $d$ is the dimension of the representation
variety in the neighborhood of smooth points. In general $d$
is different from the Zariski dimension as computed by the
cohomology $H^*(M,{\frak g}_{\theta})$. Let us define
$b(k)=\dim H^k(M,{\frak g}_{\theta})$.
Then $b(0)=b(4)=0$ by irreducibility
(and due to Poincar\'e duality), while
$b(1)=b(3)$. Recall that the space
$H^1(M,{\frak g}_{\theta})$ can be thought of as the
Zariski tangent space to ${\S}$ at
$\lbrack\theta\rbrack$; let us denote
$d_Z(\theta)=b(1)$.
So $d_Z=d$ at a smooth point, while $d_Z\ge d$ in general,
(see [GM] ).
Indeed a non-vanishing $H^2(M,{\frak g}_{\theta})$
may represent an
obstruction to the identification of the Zariski tangent space to the
smooth tangent space at the point $[\theta]$.

The Zariski dimension $d_Z$ may be computed by using the
Atiyah-Singer index theorem.
Let ${\Omega}^p({\frak g}_{\theta})$ denote the space of
all ${\frak g}_{\theta}$-valued exterior $p$-forms on $M$.
We may consider the elliptic complex
\begin{eqnarray}
0 \to \op0 \to \op1 \to \op2 \to \op3 \to \op4 \to 0
\label{tre}
\end{eqnarray}
whose cohomology is isomorphic to $H^*(M,{\frak g}_{\theta})$.
The index of the complex (\ref{tre}),
$\hbox{ind}=2d_Z(\theta)-b(2)$, may be
computed explicitly getting
$\hbox{ind}=-\dim G\chi(M)$,
so that
\begin{eqnarray}
d_Z(\theta)=-\unmezzo\dim G\chi(M)+\unmezzo b(2)
\label{quattro}
\end{eqnarray}
where the $\theta$ dependence is implicit in the twisted Betti number
$b(2)=\dim H^2(M,{\frak g}_{\theta})$.

\section{Counting minimal coverings}

It  is known that, under suitable hypotheses, the
Reidemeister
torsion can count closed (periodic) orbits of a flow on a
(hyperbolic) riemannian manifold [RS]. Our purpose is to show that
it can also count inequivalent geodesic ball coverings.\par
\vskip 0.5 cm
This result is  basically a consequence of a cardinality law
satisfied by the torsion.\par
 Let  $A$ and $B$ denote subcomplexes
of the manifold $M$, (as usual thought of as a cellular or
a simplicial complex),  with $M=A\cup B$, and let us consider a
representation
$\theta\in Hom({\pi}_1(M),G)$. Let us assume that, for every $i$,
volume
elements
${\mu}_i(A)$, ${\mu}_i(B)$, and ${\nu}_i(A)$, ${\nu}_i(B)$ are chosen
for the cochain complexes $C^i_{\frak g}(A)$,
$C^i_{\frak g}(B)$, and the corresponding cohomology groups
$H^i_{\frak g}(A)$, $H^i_{\frak g}(B)$, respectively. Let us further
assume that such volume elements determine the volume elements on
$C^i_{\frak g}(M)$ and $H^i_{\frak g}(M)$. Corresponding to this
choice
of volumes  let us
denote by ${\Delta}^{\frak g}(M|A)$,  ${\Delta}^{\frak g}(M|B)$,
and ${\Delta}^{\frak g}(M|A\cap B)$
the Reidemeister-Franz torsions associated with the subcomplexes
$A$, $B$, and $A\cap B$ respectively. Then
\begin{eqnarray}
{\Delta}^{\frak g}({H}_{A,B})
{\Delta}^{\frak g}(M|A\cup B){\Delta}^{\frak g}(M|A\cap B)=
{\Delta}^{\frak g}(M|A){\Delta}^{\frak g}(M|B)
\label{long}
\end{eqnarray}
where ${H}_{A,B}$ is the long exact cohomology sequence
associated with the short exact sequence generated by the complexes
$C_{\frak g}^*(A\cup{B})$,
$C_{\frak g}^*(A)\oplus{C_{\frak g}^*(B)}$, and
$C_{\frak g}^*(A\cap{B})$, (the correction term
${\Delta}^{\frak g}({H}_{A,B})$, associated with the twisted
cohomology groups of the above three cochain complexes, disappears
when the representation is acyclic).
\vskip 0.5 cm
In order to exploit this cardinality law,
let us consider all possible
minimal geodesic $\epsilon$-ball coverings of a manifold of bounded
geometry $M\in\Ricco$ with a given filling function
$\lambda=N^{(0)}_{\epsilon}(M)$.
Given a sufficiently small $\epsilon >0$, (in particular, smaller than
the ${\epsilon}_0$ provided by Zhu's theorem),
and
given a representation $\theta\colon{\pi}_1(M)\to{G}$, and
still denoting by $\theta$ its restrictions to representations of
the various ${\pi}_1(B_{\epsilon}(p_i))$, we can consider
the cohomologies with local coefficients in ${\frak g}_{\theta}$,
$H^*_{\frak g}({B_{\epsilon}(p_i)})$ for
$i=1,\ldots,\lambda$. We can use them as labels to
distribute over  the unlabelled balls
$\{B_{\epsilon}(p_i)\}$.
Obviously, the {\it coordinate} labelling of
the
balls arising from the centers $\{p_i\}$ are to be factored out to the
effect that the balls $\{B_{\epsilon}(p_i)\}$
are considered as a collection of
$\lambda=N^{(0)}_{\epsilon}(M)$ empty boxes over which distribute the
colours
$H^*_{\theta}({B_{\epsilon}(p_i)})$. This must be done according to
the constraint
expressed by the Mayer-Vietoris sequence, associated with the
intersection pattern of the covering, so as to
reproduce
$H^*_{\frak g}(\cup{B_{\epsilon}(p_i)})\simeq
H^*_{\frak g}(M)$.\par
\vskip 0.5 cm
\noindent We formalize these remarks as follows.\par
\vskip 0.5 cm

Let us assume that $M\in\Ricco$ has diameter $diam(M)$,
($diam(M)\leq D$), and Ricci curvature $Ric(M)\geq{r}$.
Let us consider the generic ball
${B_{\epsilon}(p_i)}\subset{M}$ as a riemannian manifold with
boundary, with metric tensor ${g_{\epsilon}(p_i)}$.
According to
the coarse-grained point of view, we can assume that  such
geodesic ball is obtained, by a re-scaling, from a corresponding ball
${\tilde{B}}_a(diam(M))$ of radius $diam(M)$ in a
space form ${\tilde M}^r_a$ of constant curvature $r$.
Notice that different balls, say $B_{\epsilon}(p_i)$ and
$B_{\epsilon}(p_k)$, with $i\not={k}$, may arise from different space
forms, resulting from different quotients of the simply connected
space of dimension $n$ and constant curvature $r$. Thus, for
$\epsilon$ sufficiently small, all balls are
locally isometric, but possibly with different underlying
topologies.\par
In particular, as far as the metric
properties are concerned,  we assume that
\begin{eqnarray}
g_{\epsilon}(p_i)=\frac{2{\epsilon}^2}{diam(M)^2}
{\tilde g}_r
\label{modello}
\end{eqnarray}
for every ball ${B_{\epsilon}(p_i)}$ with
${\epsilon}$ sufficiently small, and where ${\tilde g}_r$ denotes
the constant curvature metric on the space form
${\tilde M}^r_a$.
In terms of the filling function
$N^{(0)}_{\epsilon}(M)=\lambda(\epsilon)$, it is straightforward to
check
that (\ref{modello}) can be equivalently rewritten as
\begin{eqnarray}
g_{\epsilon}(p_i)={\rho}(M)^{-2/n}{\lambda(\epsilon)}^{-2/n}
{\tilde g}_r
\label{fillmetric}
\end{eqnarray}
where ${\rho}(M)$ is a suitable function, depending on the parameters
$n$, $r$, $D$, $V$, and on the actual geometry of the manifold
$M$.\par

\vskip 0.5 cm
For later use, it is also convenient to
introduce the
deformation parameter
\begin{eqnarray}
t(\epsilon)\equiv \ln[{\rho}(M)^{-2/n}{\lambda(\epsilon)}^{-2/n}]
\end{eqnarray}
so that we can describe the  re-scaling (\ref{modello}) as obtained
through a smooth one-parameter family of
conformal deformation
\begin{eqnarray}
g_t(p_i)=e^{t(\epsilon)}{\tilde g}_r
\label{scaling}
\end{eqnarray}
interpolating between ${\tilde g}_r$, (corresponding to
$t=0$), and the actual $g_{\epsilon}(p_i)$.
\vskip 0.5 cm

As far as topology is concerned, since the ball $B_{\epsilon}(p_i)$
comes from the re-scaling of the reference ball
${\tilde{B}}_a(diam(M))$ in the space form
${\tilde M}^r_a$, we can write
$H^*_{\frak g}(B_{\epsilon}(p_i))
\simeq{H^*_{\frak g}({\tilde{B}}_a(diam(M)))}$.
Thus, with the balls
$B_{\epsilon}(p_1),B_{\epsilon}(p_2),\ldots,
B_{\epsilon}(p_{\lambda})$
 we can associate  the
cohomology groups
$H^q_{\frak g}(B_{\bar{\epsilon}}(p_i))=
H^*_{\frak g}({\tilde{B}}_{a(i)}(diam(M)))$, where $a(i)$ labels the
possibly inequivalent space forms ${\tilde M}^r_a$
after which the balls
$\{B_{\epsilon}(p_i)\}$ are modelled. Notice that in general
$a(i)=a(k)$ for some pair $i\not={k}$ since the balls
$B_{\epsilon}(p_i)$ and $B_{\epsilon}(p_k)$ may be modelled after the
same space form ${\tilde M}^r_a$.\par
\vskip 0.5 cm
\subsection{Scaling of torsions}
\vskip 0.5 cm
At this stage, there is an important point we wish to stress, namely
that even if the twisted
cohomology of each ball is not affected by the dilation of the ball,
the corresponding volume elements in cohomology do change. In
particular, let
\begin{eqnarray}
{\bar{\mu}}_q(i)\equiv{\mu}_q({\tilde{B}}_{a(i)}(diam(M)))
\end{eqnarray}
and
\begin{eqnarray}
{\bar{\nu}}_q(i)\equiv{\nu}_q({\tilde{B}}_{a(i)}(diam(M)))
\end{eqnarray}
respectively
denote chosen (reference) volume elements
for the cochain complex
$C^q_{\frak g}({\tilde{B}}_{a(i)}(diam(M)))$,
and for the  cohomology group
$H^q_{\frak g}({\tilde{B}}_{a(i)}(diam(M)))$ associated with the
reference ball
balls ${\tilde{B}}_{a(i)}(diam(M))$ corresponding to
$B_{\epsilon}(p_i)$. The effect, on the above reference volumes, of
scaling to $\epsilon$ the radius of such ball, is described by the
following
\begin{Gaial}
Let $\lambda=N^{(0)}_{\epsilon}(M)$ denote the value of
the filling function as a function of $\epsilon$, then, as the
radius of the reference ball varies from $diam(M)$ to its
actual value,
the volume
elements
${\bar{\mu}}_q(i)$ and ${\bar{\nu}}_q(i)$ scale, as a function of
$\lambda$, according to
\begin{eqnarray}
\frac{{\nu}_q(i)}{{\mu}_q(i)}(\lambda(\epsilon))=
\frac{{\bar{\nu}}_q(i)}{{\bar{\mu}}_q(i)}
{\lambda(\epsilon)}^{-\frac{2}{n}
(q-\frac{n}{2})b(q)}
\end{eqnarray}
where the Betti number $b(q)$, (in real singular homology), is the
dimension of the
cohomology group
$H^q_{\frak g}({\tilde{B}}_{a(i)}(diam(M)))$, and where
${\mu}_q(i)$ and ${\nu}_q(i)$ respectively denote
the volume elements
for the cochain complex
$C^q_{\frak g}(B_{\epsilon}(p_i))$,
and for the  cohomology group
$H^q_{\frak g}(B_{\epsilon}(p_i))$ associated with the
given ball $B_{\epsilon}(p_i)$.
\end{Gaial}
\vskip 0.5 cm
 This result provides a basic {\it anomalous
scaling} relation satisfied by  the ratio of the volume elements
${\bar{\mu}}_q(i)$ and ${\bar{\nu}}_q(i)$ as the radius,
$diam(M)$, of the generic reference geodesic ball is shrunken to
$\epsilon$.\par
To prove this lemma  we first
evaluate $d/dt({\nu}_q(i)/{\mu_q(i)}$,
corresponding to the  deformation (\ref{scaling}),
and then integrate (in $t$) the resulting expression between $0$ and
$t$. This can be done by an obvious
extension of a construction discussed in the paper by Ray and Singer,
[RS], whereby
we proceed by considering the ratio of volume elements
$({\nu}_q(i)/{\mu}_q(i))$ as generated by a
proper choice of a base in
$C^q_{\frak g}({B_{\epsilon}(p_i)})$.\par
\vskip 0.5 cm
{\it Proof}. Let us denote by
${\cal D}^k_{\frak g}({B_{\epsilon}(p_i)})$ the space of
$C^{\infty}$-differential forms on
${B_{\epsilon}(p_i)}$ with values in the flat bundle
${\frak g}_{\theta}|{B_{\epsilon}(p_i)}$, and which satisfy
relative boundary conditions at each point of the boundary
$\partial{B_{\epsilon}(p_i)}$, (for a definition of such
boundary
conditions see Ray-Singer, {\it ibidem} p.162). Corresponding
quantities are
similarly defined also for the reference  ball
${\tilde{B}}_{a(i)}(diam(M))$.\par
\vskip 0.5 cm
Let ${\bf h}^q(t)\in {\cal H}^q$ be an orthonormal base of harmonic
$q$-forms (with coefficients in ${\frak g}_{\theta}$), in the space
${\cal H}^q\subset{{\cal D}^k_{\frak g}({B_{\epsilon}(p_i)})}$,
of
harmonic forms associated with the metric $g_t$. Let
$A^q\colon{\cal H}^q\to{C^q_{\frak g}({B_{\epsilon}(p_i)})}$
denote
the twisted deRham map
\begin{eqnarray}
A^q{\bf h}(\xi\otimes\sigma)=\int_{\sigma}tr({\xi},{\bf h})
\end{eqnarray}
where ${\sigma}$ is a $q$-cell in ${B_{\epsilon}(p_i)}$,
$\xi\in{\frak g}_{\theta}$, and $tr(\cdot,\cdot)$ denotes the inner
product in ${\frak g}_{\theta}$. Since $A^q$ is an injective map of
${\cal H}^q$ onto a linear space of cocycles representing
$H^q_{\frak g}({B_{\epsilon}(p_i)})$, we may use it as  a part
of a
base for $C^q_{\frak g}({B_{\epsilon}(p_i)})$. Choose a base
${\bf b}^q=\{b^q_j\}$ for the space of coboundaries
$B^q_{\frak g}({B_{\epsilon}(p_i)})$ and for each
$b^{q+1}_j$ take an element
${\tilde b}^{q+1}_j$ of $C^q_{\frak g}({B_{\epsilon}(p_i)})$ such
that $d{\tilde b}^{q+1}_j=b^{q+1}_j$. Both
$b^q_j$ and ${\tilde b}^{q+1}_j$ can be chosen independently of the
metric $g_t$. Thus $(b^q_j,{\tilde b}^{q+1}_j,A^q(h^q_j))$
is a base for $C^q_{\frak g}({B_{\epsilon}(p_i)})$ depending from
the metric $g_t$ only through the base of harmonic forms
$h^q_j$. Following Ray-Singer, we denote by $D^q$ the matrix
providing the trasformation from the base
${\hat \sigma}^q_j{X_k}$ of
$C^q_{\frak g}({B_{\epsilon}(p_i)})$, generated by the cells
of $C^q({B_{\epsilon}(p_i)})$ and the orthonormal base
${\frak g}_{\theta}$, and the base
$(b^q_j,{\tilde b}^{q+1}_j,A^q(h^q_j))$
introduced above. Thus
\begin{eqnarray}
\frac{{\nu}_q(i)}{{\mu}_q(i)}=
|\det D^q|
\end{eqnarray}
The  computation of the derivative of the determinant of $D^q$ is
carried out in Ray-Singer, [RS], where it is explicitly applied to the
discussion of the behavior of the Reidemeister torsion as the metric
varies, (see their Theorem 7.6). Explicitly, we get
\begin{eqnarray}
\frac{d}{dt}\ln\frac{{\nu}_q(i)}{{\mu}_q(i)}
=\sum_{j}^{b(q)}(h^q_j,\frac{d}{dt}h^q_j)_{L^2}
\label{derivative}
\end{eqnarray}
where $b(q)$ is the  Betti number, (in real singular homology), of
$H^q_{\frak g}({B_{\epsilon}(p_i)})$, and
$(\cdot,\cdot)_{L^2}$ denotes the global $L^2$-inner product in
the space of ${\frak g}_{\theta}$-twisted harmonic
$q$-forms ${\cal H}^q$, namely, for any two such forms ${\bf f}$, and
${\bf g}$,
\begin{eqnarray}
({\bf f},{\bf g})_{L^2}=
\int_{B_{\epsilon}(p_i)}tr({\bf f}\wedge{*{\bf g}})
\end{eqnarray}
where $*$ denotes the Hodge-duality operator, and $tr$ stands for
the inner product in ${\frak g}_{\theta}$.\par
\vskip 0.5 cm

The global inner product $(h^q_j,\frac{d}{dt}h^q_j)_{L^2}$ is easily
evaluated corresponding to the conformal deformation
(\ref{scaling}). Indeed, we may rewrite
$(h^q_j,\frac{d}{dt}h^q_j)_{L^2}$ as $(h^q_j,*^{-
1}\frac{d*}{dt}h^q_j)_{L^2}$, (see {\it e.g.}, proposition 6.4 of
Ray-Singer, [RS]). A straightforward computation shows that
the derivative of the Hodge map $*_t$, associated to the
$t$-flow of
metrics $g_t$, defined by (\ref{scaling}), is provided by
\begin{eqnarray}
\frac{d*_t}{dt}|_{t=0}{\bf f}= [q-\frac{1}{2}n]*{\bf f}
\end{eqnarray}
for any given $q$-form ${\bf f}$ with values in the flat bundle
${\frak g}_{\theta}$.\par
Thus
\begin{eqnarray}
(h^q_j,\frac{d}{dt}h^q_j)_{L^2}|_{t=0}=
[q-\frac{1}{2}n]
\int_{B_{{\epsilon}/2}(p_i)}tr(h^q_j\wedge{*h^q_j})=
[q-\frac{1}{2}n]
\end{eqnarray}
since the basis $h^q_j$ is orthonormal.\par
\vskip 0.5 cm
Introducing this latter expression in (\ref{derivative}) we get
\begin{eqnarray}
\frac{d}{dt}\ln\frac{{\nu}_q(i)}{{\mu}_q(i)}
=b(q)[q-\frac{1}{2}n]
\label{timederivative}
\end{eqnarray}
We integrate (\ref{timederivative}) with the initial condition
\begin{eqnarray}
\frac{{\nu}_q(i)}{{\mu}_q(i)}(t=0)=
\frac{{\bar{\nu}}_q(i)}{{\bar{\mu}}_q(i)}
\end{eqnarray}
where ${\bar{\nu}}_q(i)$ and ${\bar{\mu}}_q(i)$ respectively refer to
the original unscaled measures on the cochain complex
$C^q_{\frak g}({\tilde{B}}_{a(i)}(diam(M)))$ and on the cohomology
group
$H^q_{\frak g}({\tilde{B}}_{a(i)}(diam(M)))$.\par
With this initial condition, and if we
take into account the explicit expression of $t$ in terms of the
filling function $\lambda$ we get
\begin{eqnarray}
\frac{{\nu}_q(i)}{{\mu}_q(i)}(t(\lambda))=
\frac{{\bar{\nu}}_q(i)}{{\bar{\mu}}_q(i)}
[{\rho}^{-2/n}
{\lambda}^{-2/n}]^{(q-n/2)b(q)}
\end{eqnarray}
Thus, we eventually get
\begin{eqnarray}
\frac{{\nu}_q(i)}{{\mu}_q(i)}(\lambda(\epsilon))=
\frac{{\bar{\nu}}_q(i)}{{\bar{\mu}}_q(i)}
{\lambda(\epsilon)}^{-\frac{2}{n}
(q-\frac{n}{2})b(q)}
\end{eqnarray}
where we have traded the  term
$[{\rho}^{-2/n}]^{(q-n/2)b(q)}$ for a redefinition of the
given original unscaled measures
${\bar{\nu}}_q(i)$ and ${\bar{\mu}}_q(i)$. This completes the
proof of the stated lemma.\par
\vskip 0.5 cm
Corresponding to this re-scaling of volume elements, we can
evaluate the relation between Reidemeister torsion, for the
generic geodesic ball $B_{\epsilon}(p_i)$, as expressed
in terms of the scaled ${\nu}_q(i)$, ${\mu}_q(i)$
and unscaled measures ${\bar{\nu}}_q(i)$, ${\bar{\mu}}_q(i)$.
A straightforward computation yields
\begin{eqnarray}
{\Delta}^{\frak g}(B_{\epsilon}(p_i);{\mu}(i),{\nu}(i))=
{\Delta}^{\frak g}(B_{\epsilon}(p_i);
{\bar{\mu}}(i),{\bar{\nu}}(i))
{\lambda}^{-\frac{2}{n}
\sum_q(-1)^q(q-\frac{n}{2})b(q)}
\label{storsion}
\end{eqnarray}
Notice that the exponent of $\lambda$, {\it viz.},
${\sum_q(-1)^q(1-\frac{2}{n}q)b(q)}$ vanishes, by Poincar\'e
duality, if the ball is compact and without boundary, (in particular
it vanishes when $\epsilon\to{diam(M)}$, namely when the ball
$B_{\epsilon}(p_i)$ is expanded so as to cover the whole manifold
$M$). In this sense, it is a measure of  the presence of
the boundary. If we set
\begin{eqnarray}
{\alpha}(i)\equiv [dim(G)]^{-1}\sum_q(-1)^qqb(q)
\end{eqnarray}
and recall that $\sum_q(-1)^qb(q)=dim(G){\chi}(i)$, where
${\chi}(i)\equiv{\chi(B_{\epsilon}(p_i))}$ is the Euler-Poincar\'e
characteristic of the given ball $B_{\epsilon}(p_i)$, then we can
rewrite (\ref{storsion})as
\begin{eqnarray}
{\Delta}^{\frak g}(B_{\epsilon}(p_i);{\mu}(i),{\nu}(i))=
{\Delta}^{\frak g}(B_{\epsilon}(p_i);
{\bar{\mu}}(i),{\bar{\nu}}(i))
{\lambda}^{dim(G)\frac{2}{n}
[\frac{n}{2}{\chi}(i)-{\alpha}(i)]}
\label{anomalo}
\end{eqnarray}

\vskip 0.5 cm
\subsection{Distinct coverings in a given representation of
${\pi}_1(M)$}
\vskip 0.5 cm
With these preliminary remarks out of the way,
our strategy is to construct, out of the sequence of
$\lambda$ balls $\{B_{\epsilon}(p_i)\}$, each endowed with the metric
$g_t(p_i)$, all possible geodesic ball coverings providing and
$\epsilon$-Hausdorff approximation to the original $M$. In order to do
so, we need to consider explicitly the generalized
Meyer-Vietoris sequence associated with the covering
$\{B_{\epsilon}(p_i)\}$, (see [BT] for details). To
simplify the notation, we shall denote by $B(i)$, with
$i=1,\ldots,\lambda$, the generic open ball $B_{\epsilon}(p_i)$.
Similarly, we denote the pairwise intersections
$B(i)\cap{B(j)}$ by $B(i,j)$, triple intersections
$B(i)\cap{B(j)}\cap{B(k)}$ by $B(i,j,k)$, and so on. Recall that
for a manifold of bounded geometry, the number of mutually
intersecting balls is bounded above by a constant $d$, depending on
the parameters $n$, $r$, $D$ characterizing $\Ricco$, but otherwise
independent from $\epsilon$. Thus, independently from $\epsilon$, the
largest cluster of mutually intersecting balls which can occur for any
$M\in\Ricco$ is
$B(i_1,i_2,\ldots,i_d)$.\par
\vskip 0.5 cm
As usual [BT], we denote by $\partial_{\eta}$ the inclusion map
which ignores the $i_{\eta}$ open ball $B(i_{\eta})$ in the generic
cluster $B(i_1,\ldots,i_{\eta},\ldots)$ . For
instance
\begin{eqnarray}
\partial_i\colon\coprod_{i}B(i,j,k)\to{B(j,k)}
\end{eqnarray}
By considering the cochain complexes
$C^*_{\frak g}(B(i,j,\ldots))$ associated with the intersections
$B(i,j,\ldots)$, one can consider the restriction map
${\delta}_{\eta}$ defined by the image of the cocycles  under the
pullback map induced by the  inclusion
$\partial_{\eta}$. For instance, corresponding to the above
inclusion  we get
\begin{eqnarray}
{\delta}_i\colon{C^*_{\frak g}(B(j,k))}\to
\prod_{i}C^*_{\frak g}(B(i,j,k))
\end{eqnarray}
\vskip 0.5 cm
Thus, associated with any given minimal geodesic ball covering
$\{B(i)\}$, there is a sequence of inclusions
relating the intersections $B(i,j,k,\ldots)$ with the
packing $\coprod_{i}B(i)$
\begin{eqnarray}
\ldots\to\coprod_{i<j<k}B(i,j,k)\to\coprod_{i<j}B(i,j)
\to\coprod_{i}B(i)\to{M}
\label{inclusions}
\end{eqnarray}
and a corresponding sequence of restrictions
\begin{eqnarray}
C^*_{\frak g}(M)\to\prod_{i}{C^*_{\frak g}(B(i))}\to
\prod_{i<j}{C^*_{\frak g}(B(i,j))}\to\prod_{i<j<k}
{C^*_{\frak g}(B(i,j,k))}\to\ldots
\label{restrictions}
\end{eqnarray}
If in this latter sequence we replace the restriction maps with the
corresponding difference operator
$\delta\colon\prod
{C^*_{\frak g}(B(i_1,\ldots,i_{\eta}))}\to
\prod{C^*_{\frak g}(B(i_1,\ldots,i_{\eta},i_{\gamma}))}$ defined by
the alternating difference
${\delta}_1-{\delta}_2+\ldots(+/-){\delta}_{\eta}-
/+{\delta}_{\gamma}$, then we get the generalized Mayer-Vietoris
exact sequence
\begin{eqnarray}
0\to{C^*_{\frak g}(M)}\to\prod_{i}{C^*_{\frak g}(B(i))}\to
\prod_{i<j}{C^*_{\frak g}(B(i,j))}\to\prod_{i<j<k}
{C^*_{\frak g}(B(i,j,k))}\to\ldots
\label{GMV}
\end{eqnarray}
\vskip 0.5 cm
The sequences (\ref{inclusions}), (\ref{restrictions}),
(\ref{GMV}) intermingle the combinatorics of the geodesic ball packings and
of the corresponding coverings
with the topology of the underlying manifold $M$.\par
\vskip 0.5 cm
The function that associates with a manifold of bounded geometry
the number of distinct geodesic ball packings
extend continuously, (in the Gromov-Hausdorff topology), through
the Mayer-Vietoris sequence (\ref{GMV}). Thus, our strategy will be
to enumerate all possible ${\epsilon}/2$-geodesic ball packings,
 modulo a permutation of their centers $\{p_i\}$,
and then extend by continuity the resulting counting function to the
corresponding coverings.
\vskip 0.5 cm
In order to view a manifold of given fundamental group ${\pi}_1(M)$ as
generated by packing and gluing metric balls
we must choose base points and arcs connecting these points.
Only in this way we will be able to consider curves in the balls
either as elements of the fundamental groups of the balls themselves
or as elements of the fundamental group of the manifold $M$. So we
choose as base points in the balls their
respective $\lambda$ centers $p_1,p_2,\ldots,p_{\lambda}$. One of
these centers (say $p_1$) is then chosen as a base point in $M$.
Next we need to choose arcs $L_{ij}$ connecting the points
$p_i$ and $p_j$. This amounts in giving a {\it reference}
intersection pattern for the geodesic ball coverings,
namely a {\it reference one-skeleton}
${\Gamma}^{(1)}_{\epsilon}(M;ref)$.
If $L_i$ is a path in $M$, corresponding to a path
 in the graph ${\Gamma}^{(1)}_{\epsilon}(M;ref)$,
connecting $p_1$ with $p_i$, and $C_i$ is a curve in the ball
$B(i)$, then ${\hat C}_i\equiv L^{-1}_i*C_i*L^{-1}_i$ is an
equivalence class in ${\pi}_1(M)$. In this connection, it is
particularly helpful that isomorphic, (in the combinatorial sense),
one-skeleton graphs correspond to manifolds with a same homotopy type.
\vskip 0.5 cm

The remarks above imply  that in order to enumerate all possible
coverings, we {\it need} to start by giving a
{\it reference } covering $Cov_{ref}$

\begin{eqnarray}
\ldots\to\coprod_{i<j<k}B(i,j,k)\to\coprod_{i<j}B(i,j)
\to\coprod_{i}B(i)\to{M}
\label{refcov}
\end{eqnarray}

\noindent specifying the homotopy type of the manifolds $M$ in
$\Ricco$ we are interested in. We wish to stress that this reference
covering is common to many topologically distinct manifolds, for we
are not specifying a priori the topology of each ball. Recall that
according to
Gromov's coarse grained point of view, two manifolds $M_1$ and
$M_2$ in $\Ricco$, having
the same $\epsilon$-geodesic ball covering, define an $\epsilon$
Hausdorff approximation of a same  manifold, without
necessarily being homeomorphic to each other. For $\epsilon$ small
enough, such approximating manifolds only share the homotopy type,
(and hence have isomorphic fundamental groups). Thus the reference
covering $Cov_{ref}$, (\ref{refcov}), may be considered as a
bookkeeping device for fixing the homotopy type, (and in particular the
fundamental group), of the class of manifolds we are interested in.
\vskip 0.5 cm
{}From a combinatorial point of view, $Cov_{ref}$ labels the
intersection pattern of
centers $\{p_i\}$ of the balls in a given order. The strategy is to
determine the number of different ways of associating with such
centers the actual balls,
$\{{\tilde B}_a,H^*_{\frak g}({\tilde B}_a)\}$, after which the
geodesic balls are modelled, {\it i.e.}, we have to fill the {\it
reference} balls $B(i)$ with some topology. Any two such
correspondence between
centers and model balls are considered equivalent if they can be
obtained one from the other through the action of the symmetric group
acting on the centers. In this way we avoid to count as distinct the
re-labellings of the centers of a same pattern of
model balls. We prove
that in this way, we can obtain all possible coverings.
\vskip 0.5 cm

Let $Perm$ denote the group of permutations of the collection
of balls $\{B(i)\}\in {Cov}_{ref}$, namely the symmetric group
$S_{\lambda}$ acting on the $\lambda$ centers
$\{p_1,\ldots,p_{\lambda}\}$. Also let
$\{C^*_{\frak g}(a)\}$,  with  $a=1,2,\ldots,|C^*_{\frak g}|$,
denote the set of
possible cochain groups for
the model balls $\{{\tilde B}_a\}$, where $|C^*_{\frak g}|$ denotes
the cardinality of
$\{C^*_{\frak g}(a)\}$.\par
We are tacitly assuming that
different balls may have the same $C^*_{\frak g}(a)$.
But
actually, in the final
result we  allow for $|C^*_{\frak g}|\to (n+1)\lambda$.
As often emphasized, $\{C^*_{\frak g}(a)\}$ is the typical set of
{\it colours} for the balls $B(i)$ coming from the model balls
$\{{\tilde B}_a\}$ in
the space forms ${\tilde M}^r_a$.
 Similarly, we denote by
$\{C^*_{\frak g}(a,b)\}\subset \{C^*_{\frak g}(a)\}$ the set of
possible cochain groups for the
pairwise intersections $B(i,j)$; by
$\{C^*_{\frak g}(a,b,c)\}\subset \{C^*_{\frak g}(a)\}$ the possible
cochain groups for the
triplewise intersections $B(i,j,k)$, etc. All such groups are assumed
to be
related by a sequence of restrictions analogous to
(\ref{restrictions}), namely
\begin{eqnarray}
\prod_{a}{C^*_{\frak g}(a)}\to
\prod_{a<b}{C^*_{\frak g}(a,b)}\to\prod_{a<b<c}
{C^*_{\frak g}(a,b,c)}\to\ldots
\label{modrest}
\end{eqnarray}
\vskip 0.5 cm
\begin{Gaiar} \rm
It is  important to
stress that even if the balls, (and their intersections),  are
topologically trivial, (namely
if they are contractible), the labels associated with the
$C^*_{\frak g}(a)$ are non trivial. Indeed, for a contractible ball we
get
\begin{eqnarray}
 H^0_{\frak g}(B(i))\simeq {\frak g}_{{\theta}_i}
\end{eqnarray}

\noindent while the remaining twisted cohomology groups
all vanish. Thus in this case, the label is provided by local flat
bundles
over $B(i)$ associated with the representation $\theta$. Since there
is no canonical isomorphism between these flat bundles over
the balls $B(i)$, we have to assume that the labels
$C^*_{\frak g}(a)$ are distinct.
\end{Gaiar}
\vskip 0.5 cm
Let us consider the set  of all functions,
$f\equiv(f_{(i)},f_{(ij)},f_{(ij\ldots)},\ldots)$,
compatible with the morphisms of the two complexes
(\ref{inclusions}) and (\ref{modrest}), where
\begin{eqnarray}
f_{(i_1\ldots{i_p})}\colon\{B(i_1,\ldots,i_p)\}
\to\{C^*_{\frak g}(a_1,\ldots,a_p)\}
\end{eqnarray}
is the function which associates with the generic
mutual intersection of balls
$B(i,j,\ldots)$ the corresponding cochain group
$C^*_{\frak g}(B(i,j,\ldots))=C^*_{\frak g}(a_1,a_2,\ldots)$ out of
the possible ones
$\{C^*_{\frak g}(a,b,\ldots)\}$.
 \vskip 0.5 cm
Let $\sigma\in{Perm}$ a permutation acting on the balls $\{B(i)\}$.
Any such $\sigma$ can be made to act on the set
of function $f$, by defining
\begin{eqnarray}
({\sigma}^*f)(B(i_1,\ldots,i_k))=
f({\sigma}B(i_1,\ldots,i_k))
\end{eqnarray}
for any $1\leq{k}\leq{d}$, and where
\begin{eqnarray}
{\sigma}B(i_1,\ldots,i_k)\equiv
{\sigma}B(i_1)\cap\ldots\cap{\sigma}B(i_k)
\end{eqnarray}

Thus, the equivalence class of configurations $f$ under the
action of $Perm$ is well defined; it is the
{\it Combinatorial Pattern} of the geodesic ball covering
$f(Cov_{ref})$  in the
representation $[\theta]$.\par
\vskip 0.5 cm
Notice that
if we assume that the reference covering $Cov_{ref}$ is explicitly
realized on a given manifold $M$, then  the orbit of the map
\begin{eqnarray}
f^{(ref)}_{(i_1\ldots{i_p})}\colon\{B(i_1,\ldots,i_p)\}
\to\{C^*_{\frak g}(M;i_1,\ldots,i_p)\}
\end{eqnarray}
which allocates  the balls $\{B(i)\}$ of
the reference covering on their centers,
corresponds to the given reference covering
$Cov_{ref}$ and all  isomorphic coverings that can be
obtained from the reference covering by relabelling the centers of the
balls. Not all possible maps
$f_{(i_1\ldots{i_p})}$ belong to the orbit of
$f^{(ref)}_{(i_1\ldots{i_p})}$ and in general we can prove
the following
\begin{Gaiat}
In a given conjugacy class of (irreducible) representations
$[\theta]\in \frac{Hom({\pi}_1(M),G)}{G}$, and given a set of
possible colours (\ref{modrest}),
any two minimal geodesic  ${\epsilon}$-ball coverings
(\ref{inclusions}) are distinct
if and only if they correspond to distinct orbits of the
permutation
group $Perm$ acting on the set of all functions
$f\equiv(f_{(i)},f_{(ij)},f_{(ij\ldots)},\ldots)$
\end{Gaiat}
\vskip 0.5 cm
{\it Proof}. Let $M\in\Ricco$ be a given manifold Let $Cov_1$ and
$Cov_2$ be $\epsilon$-geodesic ball coverings of $M$ having the same
number of balls. They are {\it isomorphic}  if there
is an injective mapping $h$ of the balls of $Cov_1$ onto those of
$Cov_2$ which satisfies the following condition:\par
\noindent {\it (i)} Any two distinct balls $B_{\alpha}$ and
$B_{\beta}$ of $Cov_1$ mutually intersect each other if and only if
their images $h(B_{\alpha})$ and $h(B_{\beta})$ mutually intersects
each other in $Cov_2$.\par
This condition is extended to the mutual intersection of any number
($\leq d$), of balls, and can be rephrased in terms of the nerves
associated with the coverings $Cov_1$ and $Cov_2$, by saying that
vertices of ${\cal N}(Cov_1)$ define a $k$-simplex if and only if
their images under $h$ define a $k$-simplex in
${\cal N}(Cov_2)$, (see section 2.1). \par
\vskip 0.5 cm
Let $f^{(1)}\equiv(f_{(i)},f_{(ij)},f_{(ij\ldots)},\ldots)^{(1)}$ and
$f^{(2)}\equiv(f_{(i)},f_{(ij)},f_{(ij\ldots)},\ldots)^{(2)}$
be two functions which are in distinct orbits of the symmetric group.
Let us assume that they give rise to two isomorphic geodesic ball
coverings according to the definition recalled above. Then there is a
mapping $h$ of the balls of the covering
$f^{(1)}(Cov_{ref})$ onto the balls of the covering
$f^{(2)}(Cov_{ref})$ such that the corresponding nerves are
isomorphic. We can use the map $h$ to relabel the vertices of
$f^{(2)}(Cov_{ref})$. Thus $f^{(2)}$ and $f^{(1)}$ do necessarily
belong to the same orbit of the symmetric group, and we get a
contradiction. Conversely, let us assume that
$f^{(1)}\equiv(f_{(i)},f_{(ij)},f_{(ij\ldots)},\ldots)^{(1)}$ and
$f^{(2)}\equiv(f_{(i)},f_{(ij)},f_{(ij\ldots)},\ldots)^{(2)}$
are in the same orbit of the symmetric group. Then the permutation
which maps $f^{(1)}$ to $f^{(2)}$ is an injective mapping of the
covering defined by $f^{(1)}$ onto the covering defined by $f^{(2)}$,
and  the two coverings are  isomorphic.
\vskip 0.5 cm

Since the functions $f$ must be compatible with the
morphisms of the complexes (\ref{inclusions}) and (\ref{modrest}), and
the action of the symmetric group extends naturally through
(\ref{inclusions}), there is no need to consider all functions
$f_{(i)}$, $f_{(ij)}$, $f_{(ij\ldots)}$ as varying independently.
To generate a geodesic ball covering it suffices to assign
 the set  of all functions,
$f\equiv\{f_{(i)}\}$,
\begin{eqnarray}
f_{(i)}\colon\{B(i)\}
\to\{C^*_{\frak g}(a)\}
\end{eqnarray}
which associate with the generic ball
$B(i)$ the corresponding cochain group
$C^*_{\frak g}(B(i))=C^*_{\frak g}(a)$ out of
the possible ones
$\{C^*_{\frak g}(a)\}$. The remaining functions $f_{(ij\ldots)}$
are then determined by the given reference pattern
(\ref{inclusions}). This circumstance simply corresponds to the fact
that the assignment of a {\it geodesic ball packing}, {\it i.e.}, of
$f_{(i)}$, characterizes a corresponding geodesic ball covering, ({\it
viz.}, the one obtained by doubling the radius of the balls), and if
we estimate the number of distinct
geodesic ball packings we can also estimate the number of the
corresponding geodesic ball coverings.
\vskip 0.5 cm
Thus we need to count the number of the distinct patterns associated
with
the orbits of $f_{(i)}$ under the symmetric group. This can be
accomplished through the use of P\'olya's enumeration theorem
[Bo].\par
\vskip 0.5 cm
\subsection{Entropy function in a given representation of
${\pi}_1(M)$}
\vskip 0.5 cm
Let us write the generic permutation $\sigma\in{Perm}$ as a product of
disjoint cyclic permutations acting on the set of balls $\{B(i)\}$.
Denote by $j_k(\sigma)$ the number of cyclic permutations (cycles) of
$\sigma$ of length $k$. Recall that the {\it cycle sum} of $Perm$
is the polynomial with integer coefficients in the indeterminates
$\{t_k\}=t_1,t_2,\ldots,t_{\lambda}$ given by
\begin{eqnarray}
C(Perm;t_1,\ldots,t_{\lambda})=
\sum_{\sigma\in{Perm}}\prod_{k=1}^{\lambda}t_k^{j_k(\sigma)}.
\end{eqnarray}
Since $Perm$ is in our case the symmetric group $S_{\lambda}$
acting on $\lambda$ objects, we get
\begin{eqnarray}
C(S_{\lambda};t_1,\ldots,t_{\lambda})=
\sum\frac{{\lambda}!}{\prod_{k=1}^{\lambda}k^{j_k}j_k!}
t_1^{j_1}t_2^{j_2}\ldots{t_{\lambda}^{j_{\lambda}}}
\end{eqnarray}
where the summation is over all partitions
$j_1+2j_2+\ldots+{\lambda}j_{\lambda}=\lambda$.\par
\vskip 0.5 cm
In order to apply P\'olya's theorem we need to introduce a function
$w\colon\{C^*_{\frak g}(a)\}\to{E}$ where $E$ is an arbitrary
commutative ring. Such $w$ is meant to provide the weight of the
possible twisted cochain groups $\{C^*_{\frak g}(a)\}$. In
this way,
one can define the weight
of a configuration $f$ of such groups over the packing as
\begin{eqnarray}
w(f)=\prod_{i}w(f(B(i)))
\end{eqnarray}
Any two configurations that are equivalent under the action of
$Perm=S_{\lambda}$ have the same weight, and the weight of the pattern
associated with the $S_{\lambda}$-orbit, ${\cal O}_h$, of
a  $f$ is just $w({\cal O}_h)=w(f)$. By summing over
all distinct orbits ${\cal O}_h$, with $h=1,\ldots,l$ one gets
the pattern sum
\begin{eqnarray}
S=\sum_{h=1}^lw({\cal O}_h)
\end{eqnarray}
where ${\cal O}_1,{\cal O}_2,\ldots,{\cal O}_l$ are the distinct
patterns of the geodesic ball packings we wish to enumerate.\par
\vskip 0.5 cm
P\'olya's enumeration
theorem, (see {\it e.g.}, [Bo]), relates the above pattern
sum to the cycle sum, namely
\begin{eqnarray}
|Perm|S=C(Perm;s_1,\ldots,s_{\lambda})
\end{eqnarray}
where $|Perm|$ is the order of the group of permutations, $Perm$,
considered,
(thus, $|Perm|={\lambda}!$ in our case), and $s_k$ is the $k$-th figure
sum
\begin{eqnarray}
s_k=\sum_{a}(w(C^*_{\frak g}(a)))^k
\label{figure}
\end{eqnarray}
where the sum extends to all cochain complexes in $\{C^*_{\frak
g}(a)\}$.
\vskip 0.5 cm

Given the generic cochain $C^*_{\frak g}(a)$, a natural
candidate for
the weigth $w(C^*_{\frak g}(a))$ is its corresponding
torsion
\begin{eqnarray}
w(C^*_{\frak g}(a,\bar{\mu},\bar{\nu}))
\equiv
{\Delta}^{\frak g}(a)
\end{eqnarray}

\noindent where ${\Delta}^{\frak g}(a)$ is the Reidemeister-Franz
torsion of the cochain complex
$C^*_{\frak g}(a)$ evaluated with respect to the
unscaled reference volume elements
 $\bar{\mu}$ and $\bar{\nu}$  induced by those for the cochain complex
$C^*_{\frak g}(a)$ and the corresponding cohomology
$H^*_{\frak g}(a)$.\par
\vskip 0.5 cm
It is preferable to have these weights expressed
in terms of the reference volumes $\bar{\mu}$ and
$\bar{\nu}$ rather than the $\epsilon$-scaled
volume elements $\mu$ and $\nu$, otherwise,
according to  (\ref{anomalo}), we would get
\begin{eqnarray}
w(C^*_{\frak g}(a;\mu,\nu))
={\Delta}^{\frak g}(a;\bar{\mu},\bar{\nu})
{\lambda}^{dim(G)\frac{2}{n}
[\frac{n}{2}{\chi}(C^*_{\frak g}(a))-
{\alpha}(C^*_{\frak g}(a))]}
\end{eqnarray}

\noindent where we have set
\begin{eqnarray}
{\alpha}(C^*_{\frak g}(a))=
[dim(G)]^{-1}[\sum_q(-1)^qqb(q;a)]
\end{eqnarray}

\noindent with ${\chi}(C^*_{\frak g}(a))$ and
$b(q;a)$ respectively
denoting the Euler-Poincar\'e characteristic and the $q$-th
Betti number of $C^*_{\frak g}(a)$.\par
\vskip 0.5 cm

Such a choice for the weight  enhances the effect of the boundaries of
the balls as follows from  the presence of the anomalous scaling term
\begin{eqnarray}
{dim(G)\frac{2}{n}
[\frac{n}{2}{\chi}(C^*_{\frak g}(a))-
{\alpha}(C^*_{\frak g}(a))]}
\label{boundaries}
\end{eqnarray}

\noindent The entropic contribution
of these boundaries to the enumeration of packings can be easily seen
to be

\begin{eqnarray}
{\lambda}^{dim(G)\frac{2}{n}
[\frac{n}{2}{\chi}(C^*_{\frak g}(a))-
{\alpha}(C^*_{\frak g}(a))]\lambda}
\end{eqnarray}

\noindent thus, it is of a factorial nature, and as such quite
disturbing in controlling the thermodynamic limit of the theory.
As stressed, its origin lies in the fact that by using as reference
measures the $\epsilon$-scaled $\mu$ and $\nu$, we are implicitly
providing an intrinsic topological labelling also for the boundaries
of the balls, (indeed (\ref{boundaries}) would vanish, by
Poincar\'e duality, if the ball were closed and without boundary).
Such boundary terms are not relevant if we are interested in
coverings, and thus the weight
$w(C^*_{\frak g}(a;\mu,\nu))$
is too detailed for our enumerative purposes. The proper
choice is rather $w(C^*_{\frak g}(a;\bar{\mu},\bar{\nu}))$.
\vskip 0.5 cm
The remarks above are  an example of the
typical strategy inherent in  P\'olya's theorem. Indeed, it is
exactly the
proper choice of the weight to be associated with the colours,
that allows one to select the details of interest in the patterns
we wish to enumerate.
\vskip 0.5 cm

With these remarks out of the way, it can be easily verified
that
the weight of a configuration $f$ of the cochain
complexes $\{C^*_{\frak g}(a)\}$ over the packing
$\{B(i)\}$
is nothing but the Reidemeister torsion, in the
given representation $[\theta]$ and with respect to
the product measures $\prod_i{\bar\mu}_i$,
$\prod_i{\bar\nu}_i$, of the disjoint union
$\coprod_i\{B(i)\stackrel{f}{\rightarrow}{C^*_{\frak
g}(a)}\}$, (this is an immediate application of the
cardinality law (\ref{long}) for the torsion).

\vskip0.5 cm

We can write down  the generic $k$-th figure sum (\ref{figure}),
and a standard application of P\'olya's theorem would provide,
at least in line of principle, the required enumeration of the
distinct coverings. However, for large values of the filling
function $\lambda$  explicit  expressions are extremely difficult to
obtain. Even for small values of $\lambda$ the evaluation of the cycle
sum corresponding to the $k$-th figure sums is unwieldy
owing to the non-trivial structure of the weight we are using.\par
Nonetheless, a useful estimate of  the number of distinct covering
can be easily extracted from P\'olya's theorem. This estimate will
be  sufficient to characterize in a geometrically significant way the
rate of growth, with $\lambda$,
of the number of geodesic ball packings.\par
\vskip 0.5 cm

According to P\'olya's theorem, we get that
\begin{eqnarray}
&\sum_{h=1}^l{\Delta}^{\frak g}({\cal O}_h)=\nonumber\\
&\sum_{\sigma}
\frac{1}{J_1(\sigma)!\ldots J_{\lambda}(\sigma)!}
\left( \frac{\sum_{a}w(C^*_{\frak g}(a))}{1}
\right) ^{J_1(\sigma)}
\ldots\left( \frac{\sum_{a}
w(C^*_{\frak g}(a))^{\lambda}}
{\lambda}\right) ^{J_{\lambda}(\sigma)}
\label{pollo}
\end{eqnarray}
where the summation is over all partitions
$J_1(\sigma)+2J_2(\sigma)+\ldots+{\lambda}J_{\lambda}(\sigma)=
\lambda$.
\vskip 0.5 cm
Since we are interested in the large $\lambda$ behavior of
the above expression,
it is convenient to rewrite the figure sums in (\ref{pollo}) in a
slightly
different
way.\par
\noindent Let
${\tilde w}(C^*_{\frak g}(a))$ denote the
value of
$w(C^*_{\frak g}(a))$ corresponding to which
the torsion ${\Delta}^{\frak g}(C^*_{\frak g}(a))$
 attains its maximum over the set
of possible colours $\{C^*_{\frak g}(a)\}$, {\it viz.},

\begin{eqnarray}
{\tilde w}(C^*_{\frak g})=\max_{a}\left\{
{\Delta}^{\frak g}(C^*_{\frak g}(a)) \right\}
\end{eqnarray}
Thus, we can write

\begin{eqnarray}
s_k=\sum_{a}
w^k(C^*_{\frak g}(a;{\mu},{\nu}))
\leq
|C^*_{\frak g}|
{\tilde w}^k(C^*_{\frak g}(a;{\mu},{\nu}))
\label{figurasy}
\end{eqnarray}
where $|C^*_{\frak g}|$ denotes the number of inequivalent
cochain groups $C^*_{\frak g}(a)$
providing the possible set of colours
of the balls $B(i)$.
\vskip 0.5 cm

The generating identity determining the cycle sum for the symmetric
group is
\begin{eqnarray}
&\sum_{j=0}^{\infty}C(t_1,t_2,\ldots,t_j)u^j/j!=
&\exp\left( ut_1+u^2t_2/2+u^3t_3/3+\ldots \right)
\label{generating}
\end{eqnarray}
where $u$ is a generic indeterminate. For notational convenience,
let us set
\begin{eqnarray}
\tau\equiv
{\tilde w}(C^*_{\frak g}(a;{\mu},{\nu}))
\end{eqnarray}

If we replace in (\ref{generating}) $t_k$ with the bound,
(\ref{figurasy}), for the
figure sum $s_k$, {\it viz.},
\begin{eqnarray}
t_k=
|C^*_{\frak g}|{\tau}^k
\end{eqnarray}
then, in the sense of generating functions, we get
\begin{eqnarray}
\sum_{j=0}^{\infty}C(t_1,t_2,\ldots,t_j)u^j/j!=
\exp\left[ |C^*_{\frak g}|
(u\tau+(u\tau)^2/2+(u\tau)^3/3+\ldots) \right]=
(1-u\tau)^{-|C^*_{\frak g}|}
\end{eqnarray}

\noindent Thus

\begin{eqnarray}
C(t_1,t_2,\ldots,t_{\lambda})=
\frac{(|C^*_{\frak g}|+\lambda-1)!}{(|C^*_{\frak g}|-1)!}
{\tilde w}^{\lambda}
\end{eqnarray}
and according to P\'olya's enumeration theorem we get
that the pattern sum over all distinct orbits of the permutation
group,
acting on the $\{C^*_{\frak g}(a,b,\ldots)\}$ coloured covering
(\ref{inclusions}), is bounded by

\begin{eqnarray}
\sum_{h=1}^l{\Delta}^{\frak g}({\cal O}_h)
\leq
\frac{(|C^*_{\frak g}|+\lambda-1)!}{{\lambda}!(|C^*_{\frak g}|-1)!}
{\tilde w}^{\lambda}
\label{somma}
\end{eqnarray}

\noindent Notice that the combinatorial factor in the
above expression is
exactly
the number of $\lambda$-combinations with repetition of
$|C^*_{\frak g}|$ distinct objects.
\vskip 0.5 cm
The colour of each ball $B(i)$ has a degeneracy, (the possible
{\it shades}), equal to $n+1$, where $n$ is the dimension of the
manifold $M$. Indeed, since each ball $B(i)$ is topologically non-
trivial, its cohomology (with local coefficients)
$H^*_{\frak g}$ is generated by $n+1$, a priori distinct, groups
$H^{l}_{\frak g}$, with $l=0,1,\ldots,n$. Since there are
$\lambda$, a priori  distinct, balls we shall set in general
\begin{eqnarray}
|C^*_{\frak g}|=(n+1)\lambda;
\end{eqnarray}
with this assumption, and for  $\lambda>>1$ we get,
by applying Stirling's formula

\begin{eqnarray}
{\left[ \frac{(|C^*_{\frak g}|+\lambda-1)!}{{\lambda}!
(|C^*_{\frak g}|-1)!}\right] }_{|C^*_{\frak g}|=(n+1)\lambda }
\simeq \frac{1}{\sqrt{2\pi}}
\sqrt{\frac{n+2}{n+1}}
{\left [ \frac{(n+2)^{n+2}}{(n+1)^{n+1}}
\right ] }^{\lambda}
{\lambda}^{-\frac{1}{2}}
\left( 1+O({\lambda}^{-\frac{3}{2}}) \right)
\end{eqnarray}

\vskip 0.5 cm
It follows from the above results that the asymptotics of the
counting function, enumerating the distinct geodesic ball packings
with a torsion ${\Delta}^{\frak g}({\cal O}_h)$
in a given representation $[\theta]$, and with respect to
the product measures $\prod_i{\bar\mu}_i$,
$\prod_i{\bar\nu}_i$,  can be read off
from the bound
\begin{eqnarray}
\sum_{h=1}^l{\Delta}^{\frak g}({\cal O}_h)
\leq
\frac{1}{\sqrt{2\pi}}
\sqrt{\frac{n+2}{n+1}}
{\left [ \frac{(n+2)^{n+2}}{(n+1)^{n+1}}{\tilde w}
\right ] }^{\lambda}
{\lambda}^{-\frac{1}{2}}
\left( 1+O({\lambda}^{-\frac{3}{2}}) \right)
\label{asintoto}
\end{eqnarray}
\vskip 0.5 cm
\noindent Explicitly, let $B_{pack}({\Delta}^{\frak g};\lambda)$
denote the number of
distinct geodesic ball packings with $\lambda$ balls and with
Reidemeister torsion ${\Delta}^{\frak g}$. In terms of
$B_{pack}({\Delta}^{\frak g};\lambda)$ we can write
\begin{eqnarray}
\sum_{h=1}^l{\Delta}^{\frak g}({\cal O}_h)=
\sum_{\Delta}
B_{pack}({\Delta}^{\frak g};\lambda){\Delta}^{\frak g}
\end{eqnarray}
\vskip 0.5 cm
\noindent Since the bound (\ref{asintoto}) is a {\it fortiori} true
for each separate term appearing in the sum, we get an estimate of
the asymptotics of the
number of distinct geodesic ball packings with torsion
${\Delta}^{\frak g}$

\begin{eqnarray}
B_{pack}({\Delta}^{\frak g};\lambda)\leq
\frac{1}{\sqrt{2\pi}{\Delta}^{\frak g}({\cal O}_h)}
\sqrt{\frac{n+2}{n+1}}
{\left [ \frac{(n+2)^{n+2}}{(n+1)^{n+1}}{\tilde w}
\right ] }^{\lambda}
{\lambda}^{-\frac{1}{2}}
\left( 1+O({\lambda}^{-\frac{3}{2}}) \right)
\label{packings}
\end{eqnarray}
\vskip 0.5 cm

\noindent Rather unexpectedly, this asymptotics is
of an exponential nature, whereas one would have guessed that,
(allowing for repetitions),  there would be a factorial number of ways
of distributing $(n+1)\lambda$ distinct
labels, (the
cohomologies $H^*_{\frak g}$), over $\lambda$ {\it empty} balls.
This latter is obviously a correct guess but it does not take into
account the action of the symmetric group on the
coordinate labellings of the centers of the balls. Since we are
interested in distinct (under re-labellings) packings, we have
to factor out this action. And this reduction is responsible of the
transition from a factorial to an exponential growth
in (\ref{packings}).\par
\vskip 0.5 cm
 Another
relevant aspect of (\ref{packings})
lies  in its dependence from
the Reidemeister torsion. At this stage, this is simply a consequence
of the choice we made for the weight in applying P\'olya's theorem.
And, had we chosen $w(C^*_{\frak g}(a))=1$
we would have obtained in place of (\ref{packings}) the
estimate
\vskip 0.5 cm
\begin{eqnarray}
\frac{1}{\sqrt{2\pi}}
\sqrt{\frac{n+2}{n+1}}
{\left [ \frac{(n+2)^{n+2}}{(n+1)^{n+1}}
\right ] }^{\lambda}
{\lambda}^{-\frac{1}{2}}
\left( 1+O({\lambda}^{-\frac{3}{2}}) \right)
\end{eqnarray}
\vskip 0.5 cm
\noindent This gives a  bound to all possible
$\epsilon$-geodesic ball packings on a manifold of given fundamental
group ${\pi}_1(M)$, which is consistent with the data coming from
numerical simulations, and in dimension $n=2$ is in remarkable
agreement with the known analytical estimates, [BIZ].
\vskip 0.5 cm
The use of the torsion as weight allows for
a finer bound, where we can distinguish between different
packings, (each packing being labelled by the corresponding
torsion
${\Delta}^{\frak g}({\cal O}_h;{\bar\mu},{\bar\nu})$).  For packings
this resolution is not particularly significant,
since
the torsion of a packing does not have any distinguished topological
meaning. However,
as we pass from the geodesic ball packing
$\coprod_iB(i,\epsilon/2)$ to the corresponding covering
$\cup_iB(i,\epsilon)$, the torsion, now evaluated for the covering,
gets identified with
the torsion of the underlying manifold.
 Correspondingly, the bound
(\ref{packings}) can be extended by continuity to geodesic
ball coverings too. The explicit  passage from packings to the
corresponding coverings is an elementary
applications of Gromov's compactness for the space $\Ricco$, and we
get the following
\vskip 0.5 cm

\begin{Gaiap}
Let $M\in\Ricco$ denote a manifold of bounded geometry with
fundamental group ${\pi}_1(M)$, and let
$\theta\colon{\pi}_1(M)\to{G}$ be an irreducible representation
of ${\pi}_1(M)$ into a (semi-simple) Lie group G.
For  $\epsilon>0$ sufficiently small, let
$\{B_M(p_i,\epsilon)\}$ denote the generic minimal geodesic ball
covering
of $M$, whose  balls are labelled by the
flat bundles ${\frak g}_{\theta}(B_M(p_i,\epsilon))$
associated with the restrictions of $\theta$ to
$B_M(p_i,\epsilon)$. If we denote by
$N^{(0)}_{\epsilon}(M)\equiv\lambda$ the filling function of the
covering, then, for $\lambda>>1$, the number,
$B_{cov}({\Delta}^{\frak g};\lambda)$, of such
distinct geodesic ball coverings is bounded
above by
\vskip 0.5 cm
\begin{eqnarray}
B_{cov}({\Delta}^{\frak g};\lambda)\leq
\frac{1}{\sqrt{2\pi}{\Delta}^{\frak g}(M)}
\sqrt{\frac{n+2}{n+1}}
{\left [ \frac{(n+2)^{n+2}}{(n+1)^{n+1}}{\tilde w}
\right ] }^{\lambda}
{\lambda}^{-\frac{1}{2}}
\left( 1+O({\lambda}^{-\frac{3}{2}}) \right)
\label{asintoto2}
\end{eqnarray}
\vskip 0.5 cm
\noindent where ${\Delta}^{\frak g}(M)$ is the Reidemeister torsion
of $M$ in the representation $\theta$.
\end{Gaiap}
\vskip 0.5 cm

\noindent {\it Proof}.
{}From a combinatorial point of view,
the injection of the possible
${\epsilon}/2$-geodesic ball packings
$\coprod_i\{B(i)\}$
into the possible $\epsilon$-geodesic
ball coverings, $\cup_i\{B(i)\}$,
is a continuous map in the Gromov-Hausdorff topology.
It is also consistent with the generalized Meyer-Vietoris sequence
(\ref{GMV})
associated with the possible coverings. Thus, corresponding to
this injection
the torsions,
${\Delta}^{\frak g}({\cal O}_h)$, of the possible distinct
${\epsilon}/2$-geodesic ball packings, naturally extend to the
torsion of the underlying coverings $\cup_i\{B(i)\}$. For a
given $\lambda$, the
bound (\ref{asintoto}) depends explicitly on the topology of the
packing only through these torsions, and the set of possible packings
of
a manifold of bounded geometry is compact (it is a finite set) in the
Gromov-Hausdorff
topology. It
immediately follows, by Tiezte extension theorem, that
(\ref{asintoto})
has a continuous extension
to the counting of all inequivalent geodesic ball coverings
of the manifold $M\in\Ricco$ in the given representation $[\theta]$.
\vskip 0.5 cm
Notice that
${\Delta}^{\frak g}(M)$ plays in (\ref{asintoto2}) the role of
a normalization factor. Since there are $\lambda$ balls
$B(i)$ in $M$, and since the Reidemeister torsion is multiplicative,
${\tilde w}^{\lambda}/{\Delta}^{\frak g}(M)$ would be of the
order of 1 if the balls were disjoint, (recall that
${\tilde w}$ is the typical torsion of the generic ball). Thus,
roughly speaking, the torsion depending factor in (\ref{asintoto2})
is a measure of the {\it gluing} of the balls of the covering.
\vskip 0.5 cm
The dependence from the representation $\theta$ in
(\ref{asintoto2}) can be made more explicit.
To this end, let us assume that each ball is contractible; then
from the cardinality formula for the torsion we get that

\begin{eqnarray}
{\tilde w}={\Delta}^{\frak g}(B(i))=\sqrt{{\Delta}^{\frak g}(S^1)}
\end{eqnarray}

\noindent where ${\Delta}^{\frak g}(S^1)$ is the torsion of the
circle $S^1$ in the given representation $\theta$.
Let $A(\theta)$ be the holonomy of a generator of ${\pi}_1(S^1)$ in
the
given representation $\theta$. If the matrix $I-A(\theta)$ is
invertible,
then the flat bundle ${\frak g}_{\theta}$, restricted to the
generic ball $B(i)$, is acyclic and
\begin{eqnarray}
{\Delta}^{\frak g}(S^1)=|\det(I-A(\theta))|
\end{eqnarray}

\noindent Thus, (\ref{asintoto2}) can be written explicitly as

\begin{eqnarray}
B_{cov}({\Delta}^{\frak g};\lambda)\leq
\frac{1}{\sqrt{2\pi}{\Delta}^{\frak g}(M)}
\sqrt{\frac{n+2}{n+1}}
{\left [ \frac{(n+2)^{n+2}}{(n+1)^{n+1}}|\det(I-A(\theta))|
\right ] }^{\lambda}
{\lambda}^{-\frac{1}{2}}
\left( 1+O({\lambda}^{-\frac{3}{2}}) \right)
\label{asintoto3}
\end{eqnarray}

\section{Summing over   coverings and the volume of the
space of Riemannian structures}
\vskip 0.5 cm

The dependence from the representation $\theta$ in
(\ref{asintoto2}), and (\ref{asintoto3}) comes from the fact
that we are counting distinct coverings on a manifold $M$
endowed with a given flat bundle ${\frak g}_{\theta}$.
We can interpret this in an interesting way by saying that
(\ref{asintoto2}) is a functional associating to an
equivalence class of representations $[\theta]$, (or which is the
same,
to each flat bundle or to each gauge equivalence class of
flat connections), a statistical weight which in a sense is counting
the inequivalent riemannian structures $M$ can carry. As a matter
of fact,
from a geometric point of view,
 the term
${\tilde w}^{\lambda}/{\Delta}^{\frak g}(M)$ is related to a
measure density on the representation variety
$Hom({\pi}_1(M),G)/G$, and as such it can be used
to define an integration on
$Hom({\pi}_1(M),G)/G$. The total measure on
$Hom({\pi}_1(M),G)/G$ defined by (\ref{asintoto2}) is
the the actual entropy function for geodesic ball coverings.
By  summing this entropy function
over all $\lambda$ we get an expression that can be
considered as providing the measure of the set of all riemannian
structures of arbitrary volume and of given fundamental group.

\vskip 0.5 cm
In order to elaborate on this point, let us recall that
the torsion ${\Delta}^{\frak g}$ is a generalized
volume element in $detline(H^*_{\frak g})$. Similarly,
we may consider the product ${\tilde w}^{\lambda}$
as an element of  $detline(H^*_{\frak g})$ obtained by
pull back from $\oplus{H^*_{\frak g}(B(i))}$
to $H^*_{\frak g}(M)$ according to the Mayer-Vietoris
sequence (\ref{GMV}). Thus, the ratio
${\tilde w}^{\lambda}/{\Delta}^{\frak g}$ can be thought of
as a density to be integrated over the representation variety.
\vskip 0.5 cm

As recalled in section $3.2$, (see also [JW]),
the choice of a representation $\theta$ in the equivalence
class $[\theta]$ identifies
the twisted cohomology group
$H^1_{\frak g}$ with the Zariski tangent space
at $[\theta]$ to the representation variety
$Hom({\pi}_1(M),G)/G$. Thus, given a choice of a volume
element $\nu$ in $H^1_{\frak g}$
we may think of $\frac{{\tilde w}^{\lambda}}{{\Delta}^{\frak g}}\nu$
as providing a measure
on (the dense open  set of irreducible representations in )
$Hom({\pi}_1(M),G)/G$. \par
This construction is actually very delicate since the representation
variety $Hom({\pi}_1(M),G)/G$ is not smooth, and consequently the
density bundle may be ill-defined. The singularities come from the
reducible representations, and given a representation
$\theta$, the tangent space to the isotropy group of
such $\theta$ is $H^0_{\frak g}$, (again see section  $3.2$ or
[JW]). As already stressed we shall be ignoring the singularities of
the representation variety in the general setting. One can make an
exception for the two-dimensional case, where the structure of
$Hom({\pi}_1(M),G)/G$ is better understood.
\vskip 0.5 cm
Given the reference measure $\nu$ on
$Hom({\pi}_1(M),G)/G$,
the associated measure $\frac{{\tilde w}^{\lambda}}{{\Delta}^{\frak
g}}\nu$
is ill-behaved as $\lambda\to\infty$.
In order to take care of this problem, we introduce as a damping term
the
Gibbs factor $\exp[-a\lambda]$ which provides a discretized version of
the (exponential of the) volume of the manifold
$M$, with $a$  the (bare) cosmological constant.
In this way we have arrived at the natural setting for providing
a measure on the
space of riemannian structures of given fundamental group
 induced by the
counting function $B_{cov}({\Delta}^{\frak g};\lambda)$:\par
\vskip 0.5 cm
\begin{eqnarray}
Meas(RIEM(M),{\pi}_1(M))=\sum_{\lambda}^{\infty}
\int_{Hom({\pi}_1(M),G)/G}
B_{cov}({\Delta}^{\frak g},\lambda)
\exp[-a\lambda]d\nu([\theta])
\label{moduli}
\end{eqnarray}
\vskip 0.5 cm
This volume (\ref{moduli}) of the
corresponding space of riemannian structures
depend on the bare cosmological
constant, here in the role of a chemical potential controlling the
average number of geodesic balls. It is the partition function,
in the $\lambda\to\infty$ limit, of
a discrete model of quantum gravity based
on geodesic ball coverings, (at least when the action
can be reduced to the cosmological term). All this is strongly
reminiscent of the interplay between two-dimensional quantum Yang-
Mills theory and the intersection parings on moduli spaces of flat
connections on a two-dimensional surface, [Wt2].
In this connection it is worth stressing that the representation
variety
$Hom({\pi}_1(M),G)/G$ has a more direct geometrical meaning which
enlightens the connection with quantum gravity better than
the usual interpretation as a moduli space for flat connection. In
dimension $2$ is known that, by taking
$G\simeq PSL(2,{\Bbb R})$, the representation variety has a connected
component homeomorphic to the Teichm\"{u}ller space
of the surface. Analogous Teichm\"{u}ller components can be
characterized for other choices of the group $G$, (see {\it e.g.},
[Go], [Hi]), and thus by considering the representation variety in
place of the moduli space of complex structure as
is the case for $2D$-gravity, implies that we are considering an
extension of $2D$-gravity. In dimension larger than two, the
representation variety $Hom({\pi}_1(M),G)/G$ can be interpreted as the
deformation space  of local $G$-structures on $M$, [Go]. For instance,
if $G=O(n)$ is the orthogonal group, then
$Hom({\pi}_1(M),O(n))/O(n)$ is the moduli space of locally flat
Euclidean structures on $M$. \par
This last remark thus explains why it is natural to sum over
$Hom({\pi}_1(M),G)/G$. Indeed,
since counting coverings can be thought of as an approximation to
compute integrals over the space of riemannian structure, the  sum
over the representation variety  $Hom({\pi}_1(M),G)/G$ is needed in
order to
take into account the {\it size} of the set of metrics realizing
such $G$-structures in the space of all riemannian structure,
(of bounded geometry).

\subsection{Bounds on the critical exponents}
\vskip 0.5 cm
The relation between the counting function $B_{cov}({\Delta}^{\frak
g};\lambda)$ and the measures on the representations variety
$Hom({\pi}_1(M),G)/G$ allows us to provide bounds  on the critical
exponents associated with
$B_{cov}({\Delta}^{\frak g};\lambda)$ by  (formal)
saddle-point estimation. A sounder application of this technique would
require a deeper discussion of the properties of the
measure $\nu([\theta])$ on $Hom({\pi}_1(M),G)/G$, in particular
one needs to understand in details the extension of a measure from the
set of irreducible representations to the reducible ones
corresponding to  the singular points of the representation variety.
We are not able to address this interesting question here.
Nevertheless, we venture
since the results obtained may be helpful.
Let us fix our attention  on
the two-dimensional case first.\par
\vskip 0.5 cm

To begin with, let us be more specific on the choice of the
group $G$ into which
we are considering representations of ${\pi}_1(M)$.
A natural example is provided by $G=U(1)$. In such a case,
the $U(1)$ conjugation action on
$Hom({\pi}_1(M),U(1))/U(1)$ is trivial, and
$ Hom({\pi}_1(M),U(1))$ is just the Jacobian variety of the
riemann surface generated by the covering considered.
Moreover,
regardless of the complex structure, one has that
topologically, $ Hom({\pi}_1(M),U(1))\simeq U(1)^{2h}$
where $h$ is the genus of the surface, (see {\it e.g.},
[Go]). We can
consider the average of (\ref{asintoto}), for $n=2$, as the
representation $\theta$ runs over
$Hom({\pi}_1(M),U(1))$. Namely

\begin{eqnarray}
\frac{2}{\sqrt{6\pi}}{\left [\frac{4^4}{3^3}\right ] }^{\lambda}
\int_{Hom({\pi}_1(M),U(1))}\frac
{({\tilde w})^{\lambda}}
{{\Delta}^{\frak g}(M)}
{\lambda}^{-1/2}
\left( 1+O({\lambda}^{-3/2}) \right)d\nu([\theta])
\label{repaverage}
\end{eqnarray}

\vskip 0.5 cm
On applying Laplace method, and denoting by $Hom_0$ the
finite set in
$Hom({\pi}_1(M),U(1))$
 where the differential of $\log{\tilde w}$ vanishes
and where the corresponding Hessian is a non-degenerate quadratic
form, we can estimate the above integral in terms of
${\lambda}^{1/2}$, (which is the power of $\lambda$ characterizing the
subleading asymptotics in (\ref{asintoto})),
and obtain the bound

\begin{eqnarray}
\int_{Hom({\pi}_1(M),G)/G}
B_{Cov}({\Delta}^{\frak g},\lambda)\leq\nonumber
\end{eqnarray}

\begin{eqnarray}
\frac{2(2\pi)^h}{\sqrt{6\pi}}{\left [\frac{4^4}{3^3}\right ]
}^{\lambda}
\sum_{\theta\in Hom_0}\sqrt{a_{\theta}}
\frac
{({\tilde w}_{\theta})^{\lambda}}
{{\Delta}_{\theta}^{\frak g}(M)}
{\lambda}^{-\frac{h}{2}-\frac{1}{2}}
\left( 1+O({\lambda}^{-3/2}) \right)
\label{prima}
\end{eqnarray}
\vskip 0.5 cm
\noindent where $a_{\theta}$ is the inverse of the
determinant of the hessian
of $\log{\tilde w}$.
\vskip 0.5 cm
As recalled in the introductory remarks, we define the critical
exponent ${\eta}(G)$ associated with the entropy function
$B_{Cov}({\Delta}^{\frak g},\lambda)$
by means of the relation

\begin{eqnarray}
\int_{Hom({\pi}_1(M),G)/G}
B_{Cov}({\Delta}^{\frak g},\lambda)\equiv
Meas{\left(\frac{Hom({\pi}_1(M),G)}{G}\right) }
\exp[c\lambda]
{\lambda}^{{\eta}_{sup}-3}
\end{eqnarray}
\vskip 0.5 cm
\noindent where $c$ is a suitable constant, (depending on $G$).
Thus, corresponding to (\ref{prima}) we get the following upper bound
for the critical exponent $\eta(G)$,

\begin{eqnarray}
\eta(G=U(1)) \leq 2+\frac{1}{2}(1-h)
\label{Bcritical}
\end{eqnarray}
\vskip 0.5 cm

One may wish to compare this bound with the exact critical exponent
associated
with (\ref{Superficie}), namely

\begin{eqnarray}
{\eta}_{Sup}=2+(1-h)\left( \frac{c-25-
\sqrt{(25-c)(1-c)}}{12}\right)
\label{Scritical}
\end{eqnarray}
\vskip 0.5 cm
\noindent It follows that (\ref{Bcritical})  correctly reproduces the
KPZ scaling in the case $h=1$, (notice
however that $\eta =2$ is not a good testing ground since this value
of
the critical exponent holds for genus $h=1$ surfaces regardless both
of the presence of matter and of the fluctuations of the metric
geometry [D1]). The bound (\ref{Bcritical}) is strict both for
genus $h=0$ and $h>1$, and it remains consistent with KPZ scaling. One
may suspect that it may be also consistent with
a strong coupling of  $2D$ gravity with
matter, namely in the regime where KPZ is believed not to be reliable.
As a matter of fact, conformal field theory has not been used in
deriving our entropy estimates.
To discuss this point further,
let us
 extend the above analysis to representations
in more general groups.\par
\vskip 0.5 cm

Recall that the group $G$ is endowed with
an Ad-invariant, symmetric, nondegenerate bilinear form. This
metric induces [Go], for $n=2$, a symplectic structure on
$Hom({\pi}_1(M),G)/G$, which can be used to give meaning to
the integration of (\ref{asintoto}) over $Hom({\pi}_1(M),G)/G$,
similarly to what was done in (\ref{repaverage}).\par
\vskip 0.5 cm
More in details, if we denote by $z(\theta)$ the centralizer of
${\theta}({\pi}_1(M))$ in $G$, then the dimension of the
Zariski tangent space, $H^1_{\frak g}(M)$, to
$Hom({\pi}_1(M),G)/G$ at $\theta$ is given by [Go],[Wa]

\begin{eqnarray}
(2h-2)dim(G)+2dim(z(\theta))
\end{eqnarray}
\vskip 0.5 cm
Thus, again on formal application of Laplace method, we get the
bound (up to the usual exponential factor
$(\frac{4^4}{3^3})^{\lambda}$),

\begin{eqnarray}
\sum_{\theta\in Hom_0}(2\pi)^{(h-1)dim(G)+
dim(z(\theta))}
\sqrt{a_{\theta}}
\frac
{{\tilde w}_{\theta}^{\lambda}}
{{\Delta}_{\theta}^{\frak g}({\cal O}_h)}
{\lambda}^{[-\frac{1}{2}(h-1)dim(G)-
\frac{1}{2}dim(z(\theta))-\frac{1}{2}]}
\left( 1+\ldots \right)
\end{eqnarray}
\vskip 0.5 cm
\noindent with obvious meaning of $Hom_0$,
and where $\ldots$ stands for terms of the order
$O({\lambda}^{[-\frac{1}{2}(h-1)dim(G)-
\frac{1}{2}dim(z(\theta))-\frac{3}{2}]})$.
The corresponding bound to the critical
exponent is (for a given $\theta\in Hom_0$),

\begin{eqnarray}
{\eta}(G)\leq 2+(1-h)\frac{dim(G)}{2}+\frac{1}{2}(1-dim(z(\theta)))
\end{eqnarray}
\vskip 0.5 cm
As for the $G=U(1)$ case, the structure of this critical exponent
is consistent with KPZ scaling, and it may be a good starting point
for discussing a strong coupling regime between matter and $2D$-gravity.

\vskip 0.5 cm

\subsection{The four-dimensional case}
\vskip 0.5 cm
The four-dimensional case can be readily discussed along the same
lines of the two-dimensional case.\par
\vskip 0.5 cm
 By (formally) integrating
(\ref{asintoto}) over
$Hom({\pi}_1(M),G)/G$, and
again on applying Laplace method, we get the asymptotics
\vskip 0.5 cm
\begin{eqnarray}
\sqrt{\frac{6}{5}}
{\left [ \frac{6^6}{5^5}
\right ] }^{\lambda}
\sum_{\theta\in Hom_0}
(2\pi)^{-dim(G){\chi}(M)/4+b(2)/4-1/2}
\sqrt{a_{\theta}}
\frac
{{\tilde w}_{\theta}^{\lambda}}
{{\Delta}_{\theta}^{\frak g}({\cal O}_h)}
{\lambda}^{[dim(G)\frac{{\chi}(M)}{8}-\frac{b(2)}{8}
-\frac{1}{2}]}
\left( 1+\ldots \right)
\label{asyquattro}
\end{eqnarray}
\vskip 0.5 cm
\noindent where $\ldots$ stand for terms of the order
$O({\lambda}^{[dim(G){\chi}(M)/8-b(2)/8
-3/2]})$. As usual, $Hom_0$ denotes the
finite set in
$Hom({\pi}_1(M),G)$
 where the differential of $\log{\tilde w}$ vanishes
and where the corresponding Hessian is a non-degenerate quadratic
form. Notice that in the above asymptotics we used (\ref{quattro})
providing the formal dimension of the Zariski tangent space to
$Hom({\pi}_1(M),G)/G$. Notice also that in the above expression
we can set ${\Delta}_{\theta}^{\frak g}({\cal O}_h)=1$, (the torsion
being trivial in dimension four for a closed manifold-see the remarks
in section 3.2; the same holds in dimension two).
\vskip 0.5 cm
The bound to the critical
exponent corresponding to the estimate (\ref{asyquattro})
is (for a given $\theta\in Hom_0$),

\begin{eqnarray}
{\eta}(G)\leq \frac{5}{2}+
\frac{dim(G){\chi}(M)}{8}-\frac{b(2)}{8}
\end{eqnarray}
\vskip 0.5 cm

As recalled in the introductory remarks this bound is fully consistent
with the (limited) numerical evidence at our disposal.
In our opinion, a more careful treatment of the integration over the
representation variety may considerably improve, (also in the
two-dimensional case), these bounds. We will not address these interesting
questions any further here. In particular, one needs to understand in
considerable detail the geometry of $Hom({\pi}_1(M),G)/G$, for $n\geq
3$. For instance,
the rather naive approach to integration over the representation
variety adopted above is not suitable for the three-dimensional case.
In dimension $3$ the Reidemeister torsion is not trivial, and
integration over $Hom({\pi}_1(M),G)/G$ is rather delicate,
(see {\it e.g.}, [JW] for a remarkable analysis), and
a separate study is  needed for discussing
the three-dimensional
case in full details. A look to (\ref{asintoto}) shows that
the entropy estimates, for $n=3$, have exactly the structure
one would expect in this case. Indeed the integration of
the (Ray-Singer) torsion over a moduli space of flat connections,
(our $Hom({\pi}_1(M),G)/G$)), is the basic ingredient in
Witten's approach to $3D$-gravity [Sc]. Details on this case will be
presented in a forthcoming paper.
\vskip 1 cm
\noindent{\bf Acknowledgements}\par
\vskip 0.5 cm
One of the authors (M.C.) would like to thank J. Ambj\o rn and
B.Durhuus for many interesting discussions on entropy estimates in
quantum gravity which motivated several improvements. He is
also indebted to P.Petersen, V for a very informative discussion on
recent finiteness results in riemannian geometry.\par
This work was completed while the first author was visiting
the State University of New York at Stony Brook, and the second
was visiting the Victoria University of Wellington, New Zealand.
They thank the respective Departments of Mathematics for their
hospitality.
\par
This work was supported by the National Institute for
Nuclear Physics (INFN), by the EEC-Contract  {\it Constrained
Dynamical System}, (Human Capital and Mobility Programme)
n.~CHRX-CT93-0362, and by the research project ``Metodi
geometrici e probabilistici in fisica matematica'' of M.U.R.S.T.
\vskip 0.5 cm
\vfill\eject
\section*{References}

\begin{description}

\bibitem[Ag] {Jana3}
 M.E.Agishtein, A.A.Migdal,
Mod.Phys.Lett. {\bf A 6} (1991)1863; Nucl.Phys. {\bf B350},
(1991)690

\bibitem[Am] {Jana1}
J.Ambj\o rn, Nucl.Phys.{\bf B}(Proc.Suppl.){\bf 25A}
(1992)8;

\bibitem[ADF] {ADF}
J.Ambj\o rn, B.Durhuus, J.Fr\"{o}hlich, Nucl.Phys. {\bf B257},
(1985)433

\bibitem[AJ] {Amb}
J.Ambj\o rn, J.Jurkiewicz, {\it On the exponential bound in four
dimensional simplicial gravity}, Preprint NBI-HE-94-29 (1994).

\bibitem[BIZ] {Pari}
D.Bessin, C.Itzykson,  J.B.Zuber, Adv.Appl.Math. {\bf 1},109(1980);
E.Br\'ezin, C.Itzykson, G.Parisi, J.B.Zuber, Commun.Math.Phys.
{\bf 59}, 35 (1978); D.Bessin, Commun.Math.Phys. {\bf 69},
143 (1979);  W.J.Tutte, Canad.Journ.Math. {\bf 14}, 21 (1962).

\bibitem[Bo] {Bollobas}
B.Bollob\'as, {\it Graph theory an introductory course},
Springer Verlag, {\bf GTM 63}, New York (1979).

\bibitem[BT] {Bott-Tu}
R.Bott, L.W.Tu, {\it Differential forms in algebraic topology},
Springer Verlag, {\bf GTM 82}, New York (1982).

\bibitem[Bou] {Bou}
D.V.Boulatov, {\it On entropy of $3$-dimensional simplicial
complexes}, Preprint NBI-HE-94-37 (1994).

\bibitem[BM] {Marinari}
B.Br\"{u}gmann, E.Marinari, Phys.Rev.Lett. {\bf 70}, (1993)1908.

\bibitem[BGP] {Burago}
Yu.Burago, M.Gromov, G.Perel'man, {\it A.D.Alexandrov spaces with
curvature bounded below}, Uspekhi Mat.Nauk {\bf 47:2}, 3-51, 1992,
(Russ.Math.Surv. {\bf 47:2}, 1-58, 1992).

\bibitem[CKR] {Cat}
S.Catterall, J.Kogut, R.Renken, {\it On the absence of an exponential
bound in four dimensional simplicial gravity},
Preprint CERN-TH-7197/94 (1994).

\bibitem[CM1] {Carf}
M.Carfora, A.Marzuoli, Class.Quantum Grav. {\bf 9}(1992)595;
Phys.Rev.Lett. {\bf 62} (1989)1339.

\bibitem[CM2] {popa}
M.Carfora, A.Marzuoli {\it Finiteness theorems in Riemannian
geometry and lattice quantum gravity},
Contemporary Mathematics {\bf 132}, 171-211
(proceedings of AMS research
meeting {\it Mathematical Aspects of Classical Field Theory },
Am.Math.Soc. Eds. M.J.Gotay, J.E.Marsden, V.Moncrief (1992).

\bibitem[CM3] {IJMPA}
M.Carfora, A.Marzuoli, Intern.Journ.Modern Phys.A, {\bf 8},
1933-1980 (1993).

\bibitem[CM4] {JGP}
M.Carfora, A.Marzuoli, {\it Entropy estimates for simplicial
quantum gravity}  Preprint NSF-ITP-93-59 to appear in
Jour.~Geom.~Physics

\bibitem[Ch] {Cheeger}
 J.Cheeger {\it Critical points of distance
functions and applications to Geometry} in {\it Geometric Topology:
Recent Developments}, P.de Bartolomeis, F.Tricerri eds. Lect.Notes in
Math. {\bf 1504}, 1-38, (1991).

\bibitem[Co] {Cohe}
M.M.Cohen, {\it A course in simple homotopy theory}, GTM 10,
(Springer Verlag, New York, 1973); see also
J.W.Milnor, Bull.Amer.Math.Soc. {\bf 72} 358 (1966);
C.P.Rourke, B.J.Sanderson {\it Introduction to Piecewise-Linear
Topology}, (Springer Verlag, New York, 1982).

\bibitem[D1] {Houches}
F.David, {\it Simplicial Quantum Gravity and Random Lattices},
Lect. given at Les Houches Nato A.S.I., {\it Gravitation and
Quantizations}, (1992), Saclay Prep. T93/028

\bibitem[D2] {Dav}
F.David, Nucl.Phys. {\bf B257}, (1985)45.

\bibitem[DNF]{}
B.Doubrovin, S.Novikov, A.Fomenko {\it G\'eom\'etrie contemporaine}
(MIR, Moscou, 1987).

\bibitem[DFJ] {Durhuus}
B.Durhuus, J.Fr\"{o}hlich, T.J\'onsson, Nucl.Phys. {\bf B240},
(1984)453.

\bibitem[FRS] {Fernandez}
R.Fernandez, J.Fr\"{o}hlich, A.Sokal, {\it Random walks, critical
phenomena, and triviality in quantum field theory},TMP
(Springer-Verlag, Berlin Heidelberg 1992)

\bibitem[Fe] {Ferry}
S.C.Ferry, {\it Finiteness theorems for manifolds in Gromov-Hausdorff
space}, Preprint SUNY at Binghamton, (1993);
S.C.Ferry, {\it Counting simple homotopy types in Gromov-Hausdorff
space}, Preprint SUNY at Binghamton, (1991).

\bibitem[Fro] {Frohlich}
J.Fr\"{o}hlich,
{\it Regge calculus and discretized gravitational functional
integrals} Preprint IHES (1981), reprinted in {\it Non-perturbative
quantum field theory-Mathematical aspects and applications},
Selected Papers of J.Fr\"{o}hlich, (World Sci. Singapore 1992);

\bibitem[Go] {Goldman}
W.Goldman, Contemp.Math. {\bf 74} (1988) 169.

\bibitem[GM] {Millson}
W.M.Goldman, J.J.Millson
{\it Deformations of flat bundles over K\"{a}hler manifolds}
in Geometry and Topology eds. C.McCrory and T.Shifrin, Lect.
Notes Pure and Appl. Math. 105 M.Dekker (1987), 129-145.

\bibitem[GrP] {Pete3}
R.Greene, P.Petersen, {\it Little topology, big volume},
Duke Math.Journ., {\bf 67}, 273-290, 1992.

\bibitem[Gr1] {Grom}
M.Gromov, {\it Structures m\'etriques pour les vari\'et\'es
Riemanniennes} (Conception Edition Diffusion Information
Communication Nathan, Paris, 1981);see also
S.Gallot, D.Hulin, J.Lafontaine {\it Riemannian Geometry},
(Springer Verlag, New York,1987). A particularly clear account
of the results connected with Gromov-Hausdorff convergence of
riemannian manifolds is provided by the paper of
K.Fukaya {\it Hausdorff convergence of riemannian manifolds and its
applications}, Advanced Studies in Pure Math. $18$-I,
{\it Recent topics in differential and analytic geometry}, 143-238,
(1990).

\bibitem[Gr2]{} M.Gromov, London Math.~Soc.~Lecture Notes $182$.

\bibitem[GP] {Grov}
K.Grove, P.V.Petersen, Annals of Math. {\bf 128}, 195 (1988).

\bibitem[GPW] {Pete}
K.Grove, P.V.Petersen, J.Y.Wu, Bull.Am.Math.Soc. {\bf 20}, 181
(1989); Invent.Math. {\bf 99}, 205 (1990) (and its Erratum).

\bibitem[Hi] {Hitchin}
N.J.Hitchin, Topology {\bf 31}, 449-473 (1992).

\bibitem[JW] {Jeffrey}
L.C.Jeffrey, J.Weitsman, {\it Half density quantization of
the moduli space of flat connections and Witten's semiclassical
manifold invariants} Preprint IASSNS-HEP-91/94.

\bibitem[Ka] {Kazakov}
V.A.Kazakov, Mod. Phys.Lett.A {\bf 4},2125 (1989).

\bibitem[Ko] {Konse}
M.Kontsevitch, Commun.Math.Phys. {\bf 147} 1 (1992).

\bibitem[MSY] {Martellini}
M.Martellini, M.Spreafico, K.Yoshida, Mod.Phys.Lett. {\bf A7},
(1992)1667; see also by the same authors, {\it A generalized model for
two dimensional quantum gravity and dynamics of random surfaces for
$d>1$}, and {\it A continuous approach to $2D$ quantum gravity for
$c>1$}, Preprints (february 1994).

\bibitem[Mor] {Morrison}
K.Morrison, Contemp.Math. {\bf 74} (1988) 220.

\bibitem[Pe] {Penner}
R.Penner, Bull.Amer.Math.Soc. {\bf 15} 73 (1986);
see also: J.Harer,D.Zagier, Invent.Math. {\bf 185} 457 (1986)

\bibitem[Per] {Pereleman}
G.Perel'man, {\it Alexandrov spaces with curvature bounded below II},
Preprint LOMI, 1991.

\bibitem[Pt] {Pete2}
P.Petersen V, {\it A finiteness theorem for metric spaces},
Journal of Diff.Geom. {\bf 31}, 387-395, 1990; P.Petersen V,
{\it Gromov-Hausdorff convergence of metric spaces}, Proc. Symp.in
Pure Math., 1990 Summer Institute on Differential Geometry, {\bf 54},
Part 3, 489-504 (1993)

\bibitem[RS] {RayS}
D.B.Ray, I.M.Singer, Advances in Math. {\bf 7}(1971)145;
D.B.Ray, Advances in Math. {\bf 4}(1970)109;
D.Fried, {\it Lefschetz formulas for flows}, Contemp.Math.
{\bf 58}, Part III, 19-69, (1987);
J.Cheeger, Ann.Math.{\bf 109}(1979)259.

\bibitem[R] {Regge}
T.Regge, Il Nuovo Cimento {\bf 19}, 558 (1961);

\bibitem[Sc] {Witt}
A.Schwarz, Lett.Math.Phys.{\bf 2}(1978)247;\par
E.Witten, Commun.Math.Phys. {\bf 117}(1988)353.

\bibitem[Va] {Varsted}
S.Varsted, Nucl.Phys. {\bf B412}, (1994)406.

\bibitem[Wa] {Walker}
K.Walker, {\it An extension of Casson's invariant}, Princeton
Univ.Press, Princeton, New Jersey (1992).

\bibitem[We] {Weingarten}
D.Weingarten, Phys.Lett. {\bf 90B}, (1980)285

\bibitem[Wi] {Williams}
R.Williams, Class.Quantum Grav. {\bf 9} 1409-1422 (1992)
(for a very comprehensive bibliography and review);\par
see also H.W. Hamber, R.M. Williams, Phys.Rev.D {\bf 47}
510-532 (1993).

\bibitem[Wt1] {Witten}
E.Witten, Surveys in Diff.Geom. {\bf 1} 243-310 (1991).

\bibitem[Wt2] {Witten2}
E.Witten, Journ.~Geom.~Phys. {\bf 9} 303-368 (1992).

\bibitem[Zh] {Zhu}
S.Zhu, {\it A finiteness theorem for Ricci curvature in dimension
three}, Preprint IAS, 1992; see also S.Zhu, Bull.Amer.Math.Soc.,
Oct.1990; and S.Zhu, {\it Bounding topology by Ricci curvature in
dimension three}, Ph.D. Thesis SUNY at Stony Brook, (1990).
\end{description}
\vfill\eject
\end{document}